\documentclass[%
 reprint,superscriptaddress,
nobibnotes,
 amsmath,amssymb,
 aps,
 prb,
longbibliography,
]{revtex4-1}

\usepackage{siunitx}
\usepackage{caption}
\usepackage{subcaption}
\usepackage{bm}%
\usepackage{xcolor} 
\usepackage{graphicx}
\usepackage{float}
\usepackage[utf8]{inputenc}
\usepackage{epstopdf}
\usepackage{afterpage}
\usepackage{setspace}
\usepackage{booktabs}
\usepackage[colorlinks=true,linkcolor=black,citecolor=black]{hyperref}

\usepackage{mhchem}            
\usepackage[utf8]{inputenc}     
\usepackage[T1]{fontenc}        
\DeclareUnicodeCharacter{0308}{\"}  

\usepackage{upgreek}
\definecolor{customcolor}{RGB}{200, 0, 200}
\expandafter\ifx\csname package@font\endcsname\relax\else
 \expandafter\expandafter
 \expandafter\usepackage
 \expandafter\expandafter
 \expandafter{\csname package@font\endcsname}%
\fi
\begin{document}

\setstretch{1.0}
\title{Insights into the Structure and Dynamics of Water at Co$_3$O$_4$(001) Using a High-Dimensional Neural Network Potential}

\author{Amir Omranpour}
\email{\textcolor{black}{amir.omranpour@rub.de}}
\affiliation{Lehrstuhl f\"ur Theoretische Chemie II, Ruhr-Universit\"at Bochum, 44780 Bochum, Germany}
\affiliation{Research Center Chemical Sciences and Sustainability, Research Alliance Ruhr, 44780 Bochum, Germany}

\author{J\"{o}rg Behler}
\email{\textcolor{black}{joerg.behler@rub.de}}
\affiliation{Lehrstuhl f\"ur Theoretische Chemie II, Ruhr-Universit\"at Bochum, 44780 Bochum, Germany}
\affiliation{Research Center Chemical Sciences and Sustainability, Research Alliance Ruhr, 44780 Bochum, Germany}

\date{\today}

\begin{abstract}
Co$_3$O$_4$ is an important catalyst for the oxidation of organic molecules in the liquid phase. Still, understanding the atomistic details of Co$_3$O$_4$-water interfaces under \textit{operando} conditions remains extremely challenging. While \textit{ab initio} molecular dynamics have become an essential tool for investigating these dynamic interfaces \textit{in silico}, they are limited  to only a few picoseconds and a few hundred atoms.
In this work, we overcome these limitations by training a high-dimensional neural network potential (HDNNP) on density functional theory data, which allows us to significantly extend the accessible time and length scales. Employing this HDNNP, we perform simulations to unravel the structure, dynamics, and reactivity of Co$_3$O$_4$(001)--water interfaces in detail.
 Our simulations reveal distinct characteristics of the two possible A and B terminations. The B-terminated surface stabilizes a compact, quasi-epitaxial hydration layer with strong templating effects, enhanced hydroxylation, and a well-organized hydrogen-bond network. In contrast, the A-termination forms a more diffuse contact layer with weaker templating, lower hydroxylation, and less ordered interfacial water.
 Extended simulations further uncover proton transfer pathways, including intermittent protonation of surface hydroxyls, migration of water molecules into the epitaxial layer, and rare hydronium-like configurations.

\end{abstract}

\maketitle

\section{Introduction}\label{sec:Introduction}

Cobalt oxide (Co$_3$O$_4$) is a transition metal oxide with mixed valence states that has attracted significant interest in recent years, owing to its unique chemical, physical, and electronic characteristics. In particular, Co$_3$O$_4$ exhibits rich redox behavior due to the coexistence of Co$^{2+}$ and Co$^{3+}$ ions, remarkable catalytic activity, tunable electronic structure, and intriguing magnetic properties\cite{P6389, P7139, P7140, P7141, P6435, P6405, P6444, P6387, P7142}. These properties make Co$_3$O$_4$ highly promising for diverse technological applications, including its role as a catalyst in alcohol oxidation\cite{P6391}, water oxidation\cite{feizi2019cobalt,jiao2009nanostructured}, methane combustion\cite{hu2008selective}, and CO oxidation\cite{xie2009low}, as well as for use in lithium-ion batteries and gas sensors\cite{li2005co3o4}. Among the different structural forms, the Co$_3$O$_4$ spinel has been particularly well studied for its effectiveness in promoting oxidation reactions, notably in the selective oxidation of hydrocarbons\cite{waidhas2020secondary, hill2017site, finocchio1997ftir}.

As a catalyst, Co$_3$O$_4$ has been widely used for the oxidation of alcohols. Recently, there has been a growing interest in carrying out catalytic reactions such as oxidation of alcohols in the liquid phase\cite{P6391}, typically in aqueous solutions, which allows for milder conditions enhancing selectivity. However, for Co$_3$O$_4$ as a catalyst, the role of the aqueous environment remains unclear, as both beneficial and detrimental effects have been reported in both computational\cite{wang2012structural} and experimental\cite{P6391} studies. 

A combination of density functional theory (DFT) calculations and high-resolution scanning transmission electron microscopy (HRSTEM) was employed by Zasada et al. to study the morphology of cobalt spinel nanostructures\cite{zasada2011periodic}. Their surface energy calculations indicated the stability sequence (001) $>$ (111) $>$ (110) for the low-index facets in vacuum, which aligns with the findings of Montoya et al.\cite{montoya2011periodic} from an independent DFT investigation. Utilizing these surface energies, they performed a Wulff construction to predict the equilibrium shape of Co$_3$O$_4$ nanoparticles. The resulting model revealed rhombicuboctahedral grains predominantly exposing the (001) and (111) facets, accounting for 48\% and 41\% of the surface area, respectively, along with a smaller contribution from the (101) facet of about 11\%. These theoretical predictions are consistent with HRSTEM images of the synthesized spinel nanocrystals\cite{zasada2011periodic}.

Although extensive experimental studies\cite{schwarz2018structure,budiyanto2023impact,natarajan2021operando,tran2022understanding,qiu2024operando,wiegmann2022operando,reikowski2019operando,haunold2025hydroxylation,zhang2018probing,varhade2023crystal} have been performed on Co$_3$O$_4$--water interfaces, they often lack the spatial and temporal resolution required to investigate these complex interfaces \textit{in operando}. Thus, computational studies are crucial to achieve a comprehensive atomistic understanding. To date, investigations using static DFT calculations\cite{chen2012water,kaptagay2015water,yan2019surface,yan2019water,haunold2025hydroxylation,huo2024dft,peng2021influence} have been the dominant theoretical approach, albeit they do not explicitly account for the impact of solvation and the dynamic nature of the interface. Ab initio molecular dynamics (AIMD) simulations overcome this limitation by intrinsically incorporating dynamics and the role of finite temperatures. So far, there have been three main AIMD studies that focused purely on Co$_3$O$_4$--water interfaces: Kox et al. on (001)\cite{kox2020impact}, Gaigeot et al. on (110)\cite{creazzo2019dft}, and Kox et al. on (111)\cite{kox2024co}. As these are computationally demanding AIMD studies, all three works were limited to simulation times of about 20 picoseconds and systems consisting of a few hundred atoms. One of the principal aims of the present work is to explicitly examine the effect of time and length scales in atomistic simulations and to extend the study of Kox et al.\cite{kox2020impact} on the (001) surface to significantly larger scales. In their work, increasing water coverage enhanced Co$^{2+}$ solvation and stabilized the A-terminated surface, while for the B-termination it primarily added diffuse outer water layers without affecting the rigid first layer or the dissociation degree. However, their simulations were restricted to up to 2 monolayers (32 water molecules), thus lacking the explicit and implicit influence of bulk water. In the present study, we extend these simulations to more than 20~monolayers in a large supercell containing about 2000 water molecules, thereby allowing to describe realistically the role of solvation.

There are two possible terminations of the Co$_3$O$_4$(001) surface: the A-termination, which is predominantly terminated by Co$^{2+}$ ions, and the B-termination, which is predominantly terminated by Co$^{3+}$ ions (for more details, see Ref.~\citenum{kox2020impact}). Experimentally, determining which termination is more favorable for catalytic reactions is of utmost importance, as this holds the key to explaining why certain cation dopants, such as Fe$^{3+}$, reduce the catalytic activity of Co$_3$O$_4$~\cite{P6391}. One possible explanation is that the incorporation of Fe$^{3+}$ preferentially occurs at Co$^{3+}$ sites, which are only exposed in the B-termination, potentially altering its catalytic behavior. Therefore, a detailed characterization of these two surface terminations is a central focus of the present study.

The spatial and temporal limitations of AIMD simulations have largely been overcome by the development of machine learning potentials (MLPs)~\cite{P4885,P5673,P6102,P6112,P6131}, which have become a mainstay in atomistic simulations of condensed systems and are now extensively used to study solid--liquid interfaces\cite{P7143,schran2021machine,omranpour2024perspective,omranpour2025machine}. However, developing accurate MLPs for oxides of comparable complexity, particularly those containing transition metals in multiple oxidation states, remains a significant challenge for theoretical studies. To date, the only other examples include recent investigations on LiMn$_2$O$_4$ \cite{P5866,P5867,P6141}, hematite (Fe$_2$O$_3$) \cite{schienbein2022nanosecond}, and magnetite (Fe$_3$O$_4$) \cite{romano2025structure,romano2024structure2}, all of which employed high-dimensional neural network potentials (HDNNPs)\cite{behler2007generalized,behler2021four}. For magnetic oxides a significant complication arises from their complex spin-dependent properties, which pose challenges even for DFT, often leading to convergence issues or metastability of high-energy configurations \cite{meredig2010method}. This, in turn, complicates the construction of accurate and robust MLPs, as their quality fundamentally depends on consistent, high-quality reference data. These challenges are well documented in Ref.~\citenum{romano2025structure}, where both Gaussian Approximation Potentials (GAP)\cite{bartok2010gaussian} and HDNNPs initially struggled to produce stable potentials for the magnetite (Fe$_3$O$_4$)--water interface, primarily due to inconsistencies in the DFT+U reference dataset.

In the present work, we overcome the spatial and temporal limitations of previous \textit{ab initio} molecular dynamics studies by developing and employing a HDNNP for Co$_3$O$_4$(001)–water interfaces. This enables us to perform nanosecond-scale simulations of realistic interfacial systems comprising thousands of atoms. Using this approach, we present a detailed atomistic investigation of the Co$_3$O$_4$(001)–water interface, with particular focus on comparing the A-terminated and B-terminated surfaces. We analyze their structural ordering, interfacial water dynamics, surface hydroxylation, and dynamic surface reconstruction, thereby revealing how surface termination governs the properties of interfacial water in case of this catalytically relevant oxide.

\section{Methods}\label{sec:methods}

Machine Learning Potentials are supervised regression models that establish a mapping from atomic structures to their associated potential energies, with labeled data consisting of structures and their corresponding energies and often forces. Unlike \textit{ab initio} molecular dynamics, where energies and forces are explicitly computed on-the-fly at each timestep using electronic structure methods, MLPs such as HDNNPs provide a continuous function that represents the potential energy surface (PES). This function efficiently predicts the total energy as well as its derivatives like the forces for a given atomic configuration.

Developing an MLP involves two essential tasks: representation and regression. The representation problem is addressed by employing descriptors that encode the local atomic environments, transforming Cartesian coordinates into features that are invariant under translation, rotation, and permutation of identical atoms. In the case of HDNNPs, this is typically achieved through atom-centered symmetry functions (ACSF)~\cite{behler2011atom}, which act as structural fingerprints to ensure that equivalent configurations yield identical energy predictions. The regression task is then performed by training a multilayer feed-forward neural network to learn the relationship between these descriptors and the target energies (and optionally forces), thus constructing an accurate model of the PES.

In HDNNPs, the local chemical environment of an atom, defined by all neighboring atoms within a cutoff sphere of radius $R_{\mathrm{c}}$, is encoded by a vector of ACSF values. In second-generation HDNNPs~\cite{behler2007generalized}, the cutoff $R_{\mathrm{c}}$ must be sufficiently large to capture all relevant interactions, and typically cutoffs between 5 and 10~\AA{} are employed. In this work, two types of ACSFs are used: radial and angular symmetry functions, as defined in Ref.~\citenum{behler2011atom}. Depending on the complexity of the system, between 30 and 150 ACSFs per atom are typically employed.

Beyond specifying the chemical elements of the atoms, no additional information such as atom types, fixed oxidation states, or predefined bonds is required. This makes HDNNPs inherently reactive, capable of accurately describing bond formation and breaking as well as changes in oxidation states, all governed by the underlying electronic structure. Since the dimensionality of the ACSF vectors is determined by the chosen set of symmetry functions and is independent of the particular local environment, these vectors can serve directly as input vectors to atomic neural networks that require a fixed dimensionality.

A separate atomic neural network is constructed for each chemical element \(\alpha\), which processes the local structural environment of each atom \(n\) to produce its atomic energy contribution \(E_n^\alpha\). The total energy \(E\) of the system, comprising \(N_\text{elements}\) elements and \(N_\text{atoms}^\alpha\) atoms of element \(\alpha\), is then given by

\begin{equation}
E = \sum_{\alpha=1}^{N_\text{elements}} \sum_{n=1}^{N_\text{atoms}^\alpha} E_n^\alpha.
\end{equation}

These atomic contributions are predicted by multilayer feed-forward neural networks, which use a linear activation function in the output layer and, in most cases, hyperbolic tangent functions in the hidden layers. For each atom, the ACSF vector is computed and passed through the appropriate neural network, with the resulting atomic energies summed to yield the total potential energy of the system. Forces, which can also be used for training, can then be obtained as analytical derivatives of the energy with respect to the atomic positions.

The weights of all atomic neural networks are optimized simultaneously using iterative gradient-based procedures. In this study, an adaptive, global, extended Kalman filter~\cite{kalman1960new,blank1994adaptive} is employed to minimize the errors of both total energies and atomic force components for a DFT training set of reference structures. For further details on the construction, training, and applications of HDNNPs, the reader is referred to several comprehensive reviews~\cite{behler2021four,behler2017first,behler2014representing,behler2015constructing,tokita2023train,omranpour2024perspective,omranpour2025machine}.

\section{Computational Details}\label{sec:Computational}

\subsection{Density Functional Theory Calculations}

All electronic structure calculations were performed using the Vienna \textit{Ab initio} Simulation Package (VASP)~\cite{kresse1996efficient,kresse1996efficiency} version~6.3.2, employing spin-polarized DFT. Exchange--correlation effects and van der Waals interactions were treated using the optPBE-vdW functional~\cite{perdew1996generalized,klimevs2009chemical,klimevs2011van}.
On-site Coulomb interactions were further accounted for via the DFT+U approach to properly treat the strongly correlated Co \(3d\) electrons, with an effective Hubbard U parameter of 2.43~eV determined following Dudarev et al.\cite{dudarev1998electron}.

Ionic cores were represented using Projector Augmented Wave (PAW) potentials~\cite{blochl1994projector}, as formulated by Kresse and Joubert~\cite{kresse1999ultrasoft,kresse1996efficiency}. Plane-wave expansions were performed with a kinetic energy cutoff of 500~eV. The Brillouin zone was sampled using a Monkhorst-Pack scheme with a \(5 \times 5 \times 5\) k-point mesh for bulk Co$_3$O$_4$, a \(3 \times 3 \times 3\) k-point mesh for bulk water, and a \(3 \times 3 \times 1\) k-point mesh for the Co$_3$O$_4$--water interfaces. The typical bulk Co$_3$O$_4$ systems consisted of 56 atoms, with a simulation box of about \(8\,\text{\AA} \times 8\,\text{\AA} \times 8\,\text{\AA}\) (see Ref.~\cite{omranpour2025machine}). The typical systems for bulk liquid water consisted of 192 atoms, with a simulation box of about \(12.42\,\text{\AA} \times 12.42\,\text{\AA} \times 12.42\,\text{\AA}\), in addition to some ice structures and liquid--vacuum interfaces (see Ref.~\citenum{schran2020committee}). The interface systems consisted of 176 atoms, and the typical simulation box dimensions were about \(8\,\text{\AA} \times 8\,\text{\AA} \times 20\,\text{\AA}\). Gaussian smearing of 0.1~eV was applied to determine partial occupancies, and non-spherical contributions within the PAW spheres were included. Electronic self-consistency was considered achieved when total energy differences fell below \(10^{-6}\)~eV.

\subsection{Construction of the Reference Data Set}

When constructing the reference data for systems such as Co$_3$O$_4$–water interfaces, it is necessary to sample three regions of configuration space corresponding to three types of atomic environments: bulk Co$_3$O$_4$, bulk water, and the Co$_3$O$_4$–water interfaces~\cite{omranpour2025machine}. For the bulk Co$_3$O$_4$ reference data, the set of 10,700 structures from Ref.\citenum{omranpour2024high} was adopted without modification. This dataset spans temperatures from 0~K up to 700~K, thus exceeding the temperature range typically used in catalysis at solid-liquid interfaces. For bulk water, 814 structures from Ref.~\citenum{schran2020committee} were employed, and energies and forces were recomputed using the electronic structure settings of this work. This particular water dataset was chosen for its efficiency and diversity, as it was curated from a large pool of aqueous configurations to accurately reproduce the key properties of water.

To build an initial dataset for Co$_3$O$_4$--water interfaces for training a preliminary set of HDNNPs, AIMD simulations were conducted with VASP on a single unit cell of Co$_3$O$_4$ exposed to 40 water molecules, with the (001) surfaces consisting of both A- and B-terminations in contact with water. The electronic structure settings mirrored those described earlier, except that a \(\Gamma\)-centered \(1 \times 1 \times 1\) k-point grid was used to enhance computational efficiency. Simulations were performed in the $NVT$ and $NPT$ ensembles at temperatures ranging from 300~K to 600~K to generate diverse interfacial structures, using a Nosé–Hoover thermostat to regulate temperature~\cite{P2758}. Each simulation ran for approximately 15~ps with a time step of 0.5~fs. The initial 4000 configurations (corresponding to about 2~ps) were discarded to allow for equilibration. Subsequently, configurations were extracted every 50 steps, yielding an initial dataset of roughly 3000 structures. For these configurations, single-point electronic structure calculations were performed with a dense \(3 \times 3 \times 1\) k-point mesh to obtain precise energies and forces.

Following the creation of this dataset and training of initial HDNNPs, an active learning strategy, following Eckhoff et al.\ \cite{eckhoff2021high,eckhoff2019molecular}, was employed to identify additional structures associated with high prediction uncertainty across an ensemble of similarly trained HDNNPs. The HDNNPs differed by random seeds, resulting in varied train/test partitions and different weight initializations. Electronic structure calculations were then performed on these high-uncertainty structures, which were added to the dataset to further improve the HDNNPs. This active learning cycle was repeated multiple times until the variance across all sampled structures was comparable to the RMSEs of the models~\cite{tokita2023train}.

\subsection{Construction of the High-Dimensional Neural Network Potential} 

The HDNNPs were constructed using the RuNNer code~\cite{behler2015constructing,behler2017first} (version dated August 16, 2023). The atomic neural networks for each element employed two hidden layers with 25 and 20 neurons, respectively. The input nodes corresponded to the element-specific ACSFs, and a single output node provided the atomic energy. A cutoff radius of $R_c = 12\,a_0$ (6.35~\AA) was used for the ACSFs to ensure an accurate description of atomic interactions in the Co$_3$O$_4$–water interfaces. The Kalman filter parameters were set to $\lambda = 0.98000$ and $\nu = 0.99870$. Further details on the employed symmetry functions and their parameters, the RuNNer settings for constructing the HDNNP, as well as energy and force correlation plots, are provided in the supporting information.

After completing the iterative generation of the reference dataset via active learning, a total of 3540 Co$_3$O$_4$(001)–water interface structures (each containing 176 atoms) were obtained. Combined with the previously generated 10,766 bulk Co$_3$O$_4$ structures~\cite{omranpour2024high} and 812 bulk water structures~\cite{schran2020committee}, this resulted in a dataset of 15,118 structures, providing 15,118 potential energies and their associated atomic force components. Of these, 90\% were used for training the HDNNP and 10\% for testing. In each epoch 100\% of all energies and 5\% of all forces were used in the training process. The root mean squared error (RMSE) for the energy is 1.201 meV/atom for the training set and 1.288 meV/atom for the test set, while the RMSE for the force components is 0.1277 eV/\AA{} and 0.1284 eV/\AA{}, respectively, which are typical for state-of-the-art MLPs. 

\subsection{Molecular Dynamics Simulations}

\begin{figure}[H]
\centering
\includegraphics[width=0.4\textwidth, trim= 480 0 460 0, clip=true]{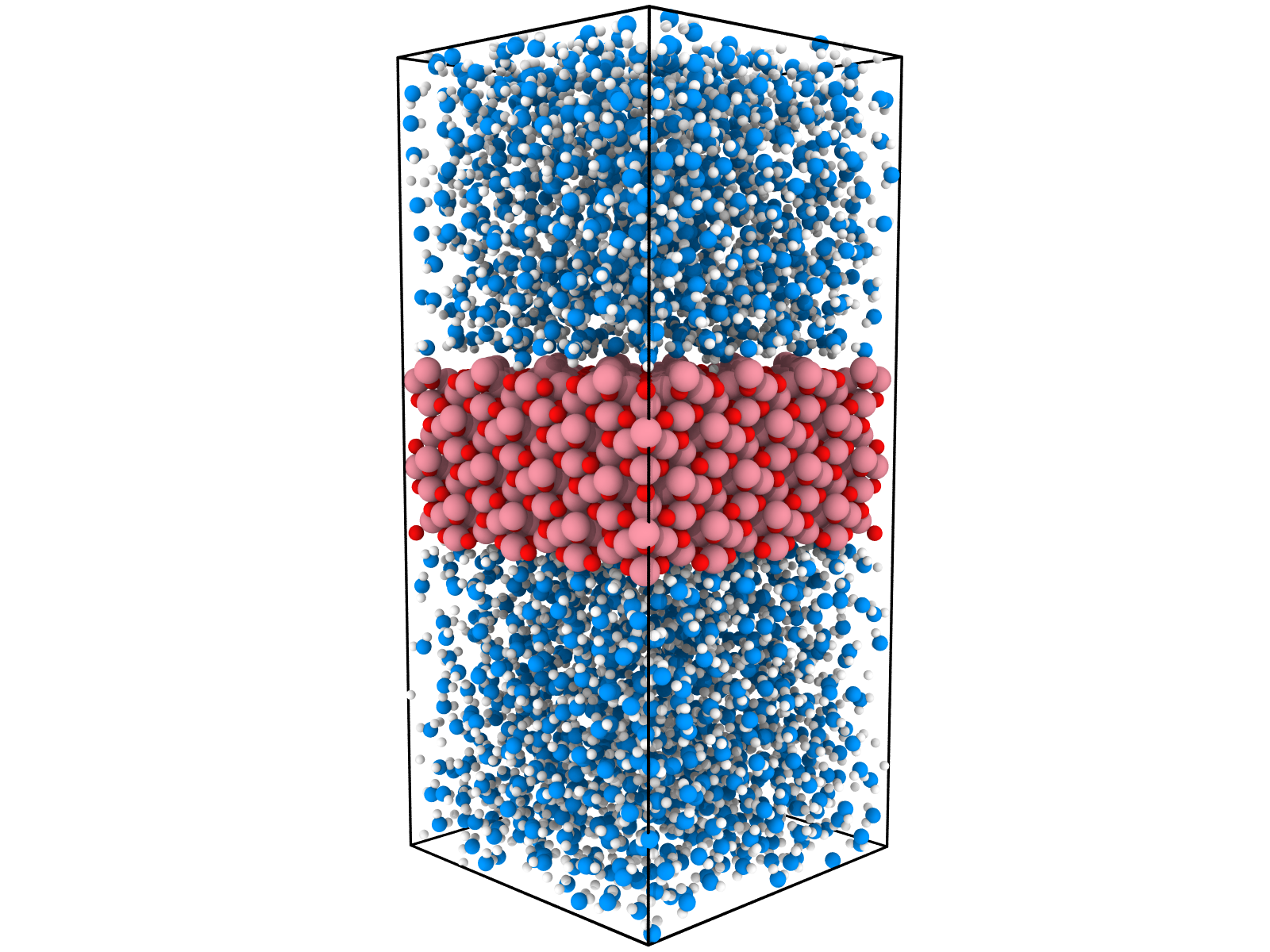}
\caption{Simulation cell consisting of a Co$_3$O$_4$(001) slab in contact with liquid water. The system contains a total of 7,936 atoms. The lateral dimensions of the simulation box are approximately 33~\AA{} in the $x$ and $y$ directions, and about 75~\AA{} in the $z$ direction. The Co$_3$O$_4$ slab, with a thickness of ~17~\AA{}, corresponds to a \(4 \times 4 \times 2\) supercell. Cobalt atoms are shown in pink, oxygen atoms in the Co$_3$O$_4$ slab in red, oxygen atoms in water in blue, and hydrogen atoms in white.}
\label{fig:cell}
\end{figure}

Molecular dynamics simulations were carried out using the Large-scale Atomic/Molecular Massively Parallel Simulator (LAMMPS)~\cite{plimpton1995fast} (version from 2$^{\mathrm{nd}}$ August 2023), incorporating the n2p2 library (version 2.2.0 from 23$^\mathrm{rd}$ May 2022)~\cite{singraber2019parallel} for employing HDNNPs. Simulations used for active learning~\cite{eckhoff2021high,eckhoff2019molecular} were conducted in the $NVT$ and $NPT$ ensembles in a temperature range from 300 to 450~K, with a time step of $\Delta t = 0.5 \, \text{fs}$. The velocity Verlet scheme~\cite{swope1982computer} was used as the integration algorithm. Temperature and pressure were controlled using the Nosé--Hoover thermostat and barostat~\cite{nose1984molecular,hoover1985canonical,P2758}, with all $NPT$ simulations performed at $p = 1$~bar. For the final simulations, systems were first equilibrated in the $NPT$ ensemble for a 1~ns run at 300~K and 400~K, and the resulting configurations were then used to perform production runs in the $NVT$ ensemble at 300~K and 400~K, respectively, for an additional 1~ns. The same simulation setup was previously applied to the ZnO--water interface using HDNNPs~\cite{quaranta2017proton}.
Figure~\ref{fig:cell} shows the \(4 \times 4 \times 2\) Co$_3$O$_4$ supercell representing the slab in the MD simulations, with the system comprising a total of 7,936 atoms. The top surface of the slab corresponds to the A-termination, while the bottom surface corresponds to the B-termination.

\section{Results and Discussion}\label{sec:results}
The structure of this section is as follows: first, a few of the most important reaction mechanisms near the interfaces are discussed in detail (Section~\ref{sec:Mechanism}). This is followed by an analysis of the contact layer structure, where snapshots of different contact layers and radial distribution functions are provided to characterize the water structure near the surfaces (Section~\ref{sec:Contact_Layer}). Subsequently, we discuss the density profiles of O and H atoms to describe the overall water distribution throughout the system (Section~\ref{sec:Density Profiles}). Thus, the section progresses from a detailed, zoomed-in view of the reaction mechanisms to the contact layers and finally to the entire system. In the end, we discuss the implications of our findings for catalysis (Section~\ref{sec:implications}).

\subsection{Proton Transfer Mechanism}\label{sec:Mechanism}

\begin{figure*}[ht]
\centering
\begin{subfigure}{0.48\textwidth}
    \centering
    \caption{}
    \label{fig:PT-1}
    \includegraphics[width=\textwidth, trim=0 0 0 0, clip=true]{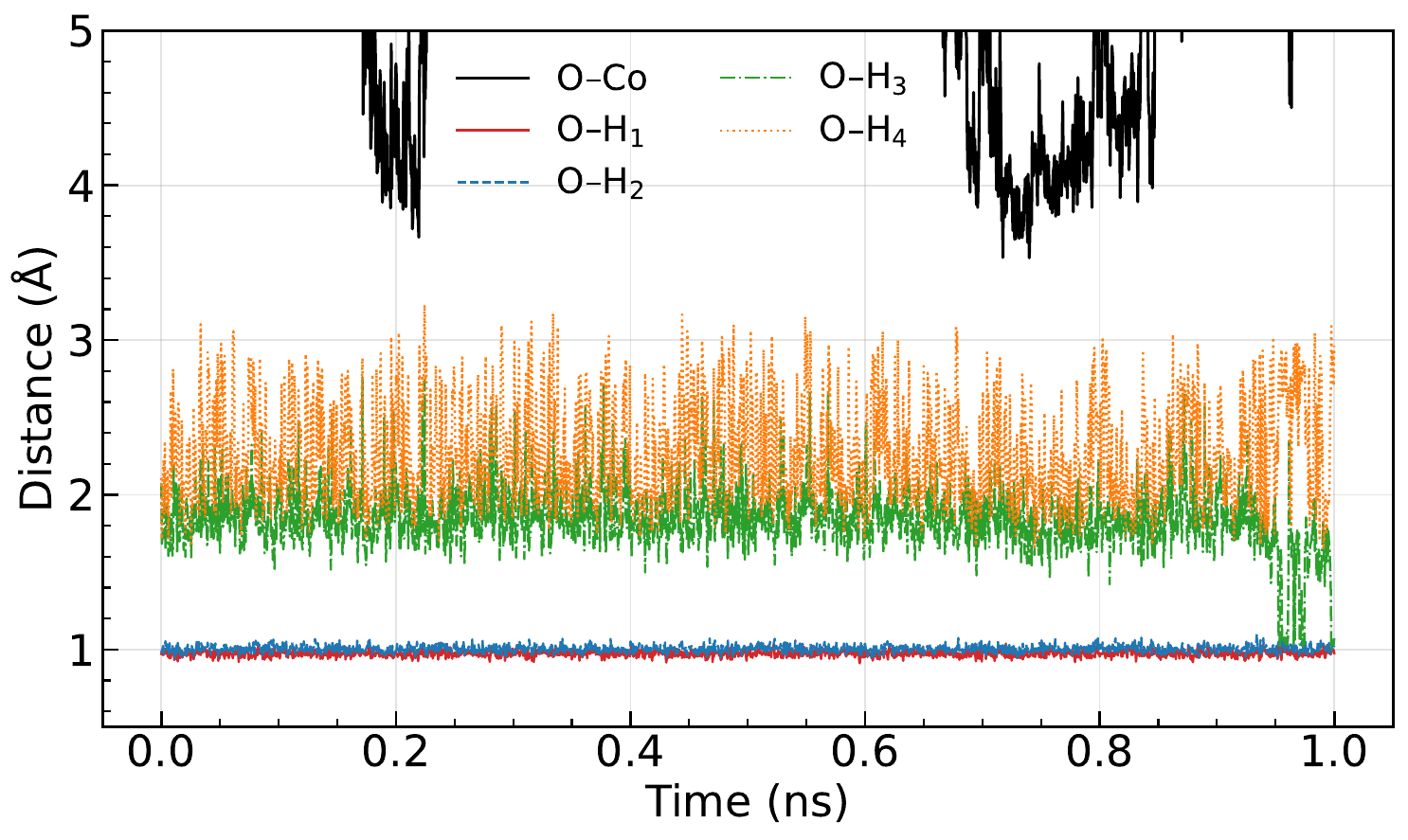}
\end{subfigure}%
\hfill
\begin{subfigure}{0.48\textwidth}
    \centering
    \caption{}
    \label{fig:PT-2}
    \includegraphics[width=\textwidth, trim=0 0 0 0, clip=true]{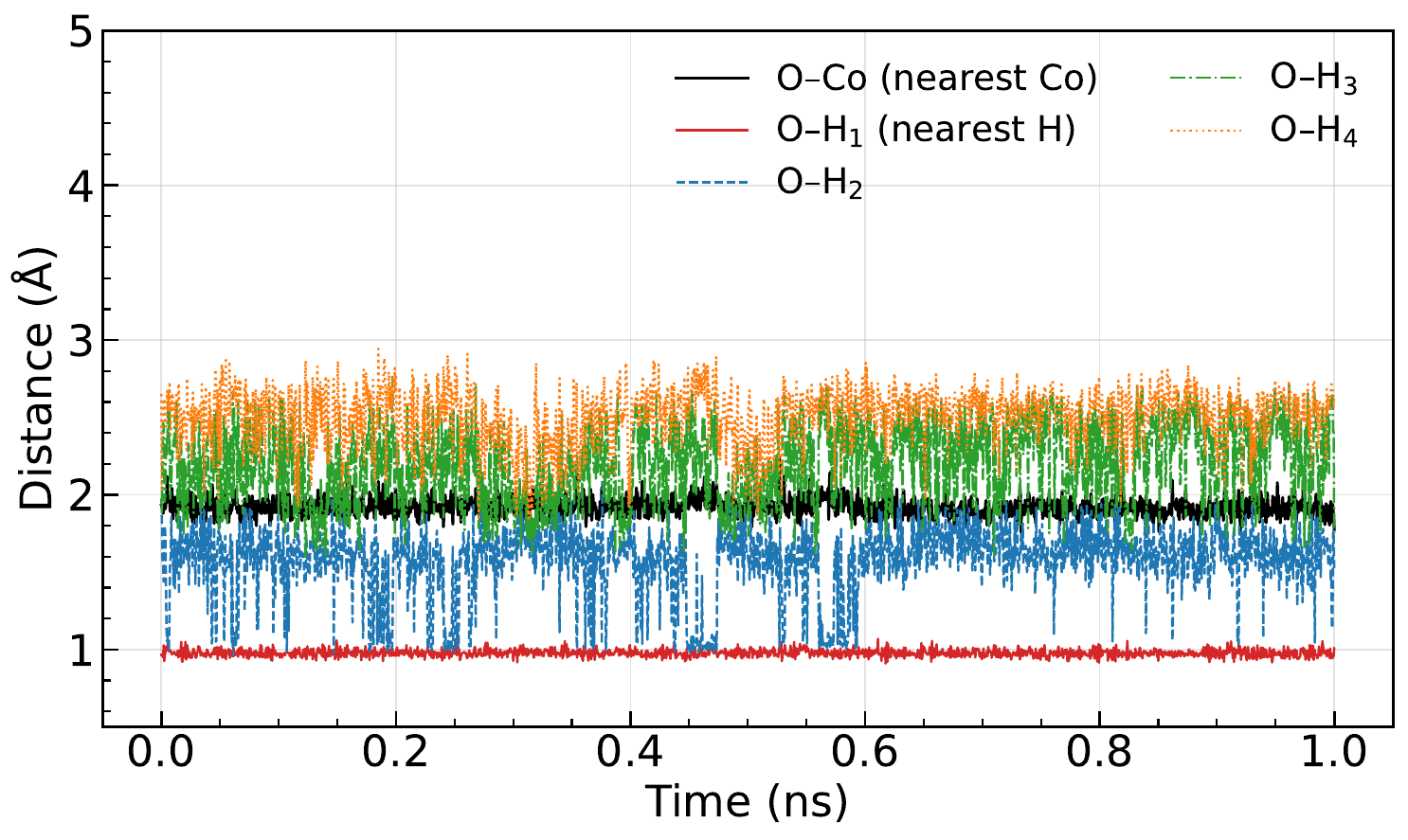}
\end{subfigure}
\vspace{0.4cm} 
\begin{subfigure}{0.48\textwidth}
    \centering
    \caption{}
    \label{fig:PT-3}
    \includegraphics[width=\textwidth, trim=0 0 0 0, clip=true]{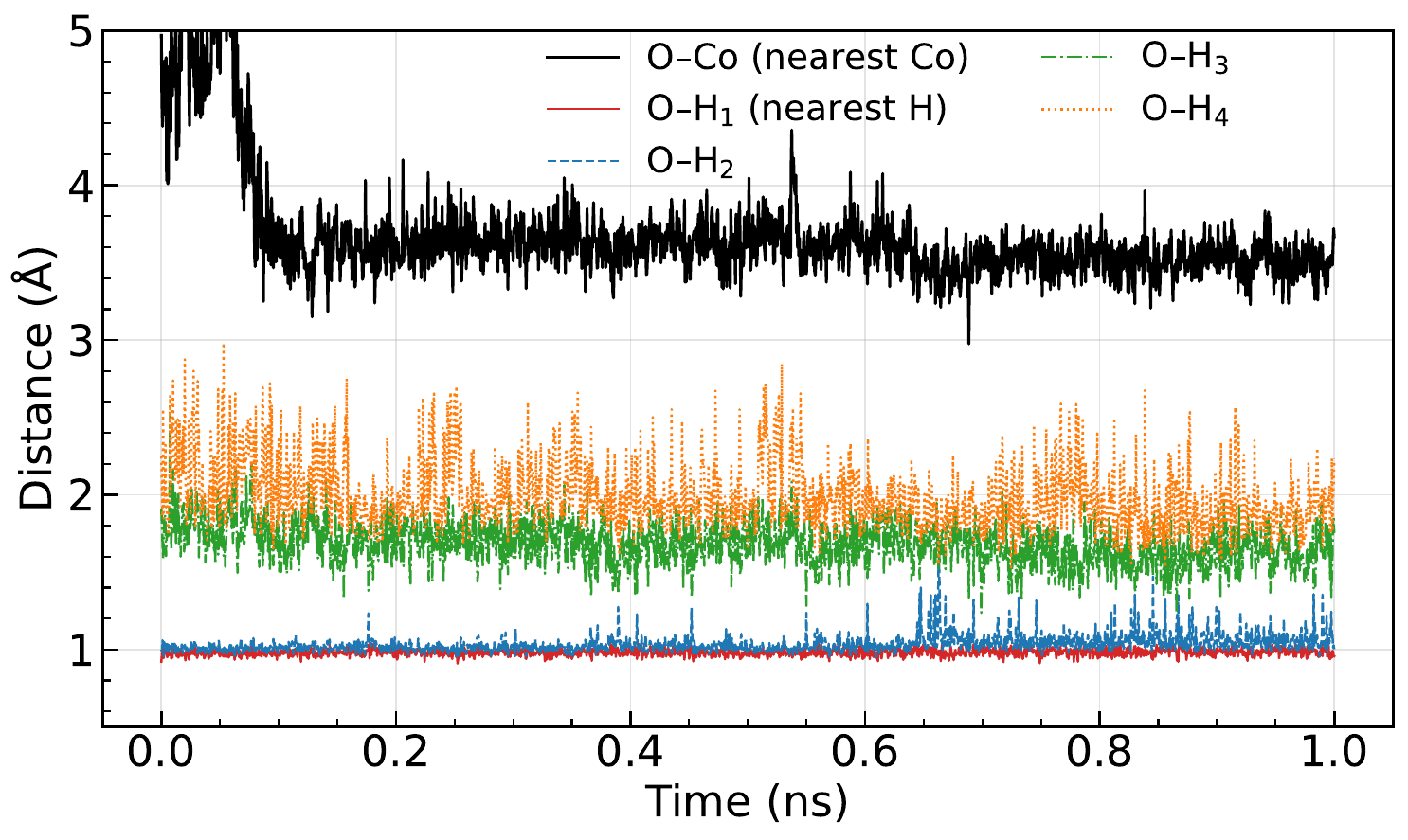}
\end{subfigure}%
\hfill
\begin{subfigure}{0.48\textwidth}
    \centering
    \caption{}
    \label{fig:PT-4}
    \includegraphics[width=\textwidth, trim=0 0 0 0, clip=true]{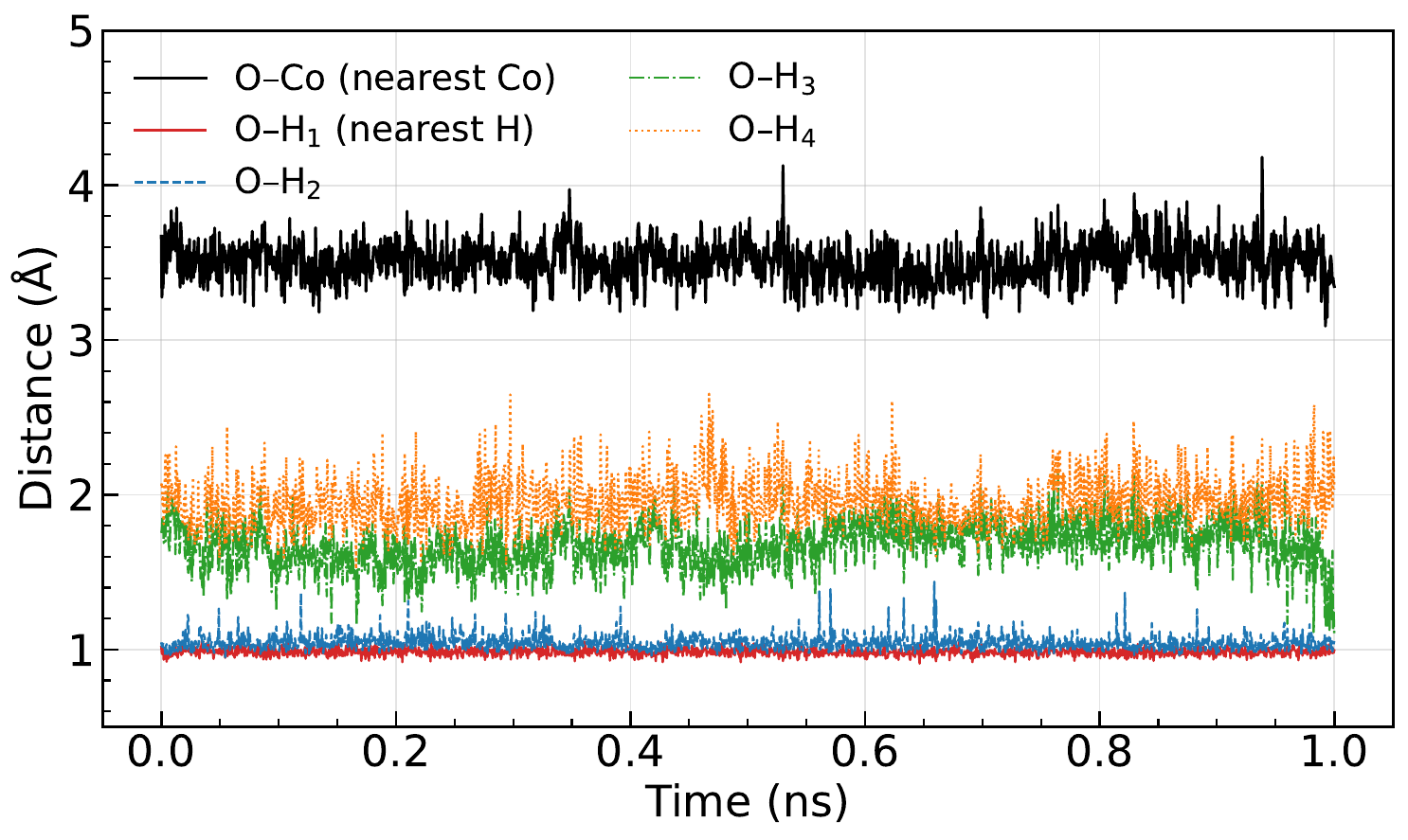}
\end{subfigure}
\caption{Proton transfer and interfacial dynamics at the Co$_3$O$_4$(001)--water interface, illustrated via time series of O--H and O--Co distances for four representative cases. 
(a) \emph{Outer hydration layer}: the tagged O starts at about $10~\text{\AA}$ from the A-terminated surface and approaches as close as $\approx4~\text{\AA}$ (black trace), while during the brief hydronium-like episode its distance is $\approx5~\text{\AA}$, indicating that it remains under the influence of the surface. The molecule maintains two persistent O--H bonds near $\approx1.0~\text{\AA}$ and two hydrogen-bond donor distances around $1.6\!-\!2.2~\text{\AA}$. A transient hydronium-like event occurs when a third H briefly approaches $\approx1.0~\text{\AA}$.  
(b) \emph{Chemisorbed hydroxyl}: the O--Co distance remains at $\approx2.0~\text{\AA}$, identifying an inner-sphere OH bound to surface Co; intermittent protonation/deprotonation events convert the OH into H$_2$O and back.  
(c) \emph{Water migrating to the epitaxial layer}: the tagged molecule initially resides in the outer hydration region with an O--Co distance of $\approx4.5\!-\!5.0~\text{\AA}$ during the first $\sim100$~ps. It then relocates to the epitaxial (outer-sphere) layer, where the O--Co distance stabilizes at $\approx3.3\!-\!3.6~\text{\AA}$. This position places it in close interfacial proximity but still distinct from chemisorbed species in panel~b at $\approx2.0~\text{\AA}$. Strengthened hydrogen bonding emerges after migration, reflecting enhanced interfacial exchange.  
(d) \emph{Outer-sphere water with late protonation}: a tagged water in the epitaxial layer undergoes a late hydronium-like transition, where a third O--H bond forms while the fourth H recedes.}
\label{fig:PT}
\end{figure*}

In liquid water and at oxide--water interfaces, protons can migrate via proton transfer events (i) in the bulk liquid through the hydrogen-bond network or (ii) near the surface through reactions involving adsorbed species (e.g., hydroxyls) and/or lattice oxygen atoms. To visualize such events in our trajectories, we monitor, for selected oxygen atoms (the “tagged” O), the time series of its distances to the nearest four hydrogen atoms (H\(_1\)–H\(_4\)) and, when relevant, to the nearest surface Co atom. In these plots, two persistent O--H distances near \(\approx1.0~\text{\AA}\) indicate a neutral water molecule; one near \(1.0~\text{\AA}\) indicates a hydroxyl; and three near \(1.0~\text{\AA}\) sustained over multiple frames is the hallmark of a hydronium-like configuration. H\(_3\)/H\(_4\) values in the \(1.5\!-\!2.4~\text{\AA}\) range report non-bonded, hydrogen-bond donor hydrogens. An O--Co distance near \(2.0~\text{\AA}\) signals an inner-sphere Co--O bond (chemisorbed OH or H\(_2\)O), whereas a distance of \(3.2\!-\!3.6~\text{\AA}\) is typical for outer-sphere water in the epitaxial/solvation layer.

\paragraph{Outer hydration layer case (A-termination).}
Figure~\ref{fig:PT}a shows a tagged water molecule located relatively far from the surface. Its initial distance from the A-terminated surface is about \(10~\text{\AA}\), and the closest it approaches is roughly \(4~\text{\AA}\), as indicated by the black trace in the figure. During the brief hydronium-like episode discussed below, its distance from the surface is around \(5~\text{\AA}\), implying that even at this range, the molecule remains under the influence of the surface. The red and blue traces (H\(_1\), H\(_2\)) stay close to \(1.0~\text{\AA}\) throughout, confirming two covalent O--H bonds and no persistent dissociation. The green/orange traces (H\(_3\), H\(_4\)) fluctuate around \(1.6\!-\!2.2~\text{\AA}\), consistent with typical hydrogen-bond donors in the first and second hydration shells. Near the end of the trajectory, H\(_3\) transiently approaches \(\sim1.0~\text{\AA}\) while H\(_4\) simultaneously recedes toward \(\sim2.5\!-\!2.7~\text{\AA}\). This concerted behavior corresponds to a short-lived hydronium-like configuration in which the tagged O momentarily binds three protons before rapidly relaxing back to water. In this example, the residence time is on the order of \(\sim 50\)~ps, underscoring that protonation events in classical MD can be rare and require long sampling to capture convincingly.

\paragraph{Chemisorbed hydroxyl on Co (B-termination).}
In Fig.~\ref{fig:PT}b the black O--Co curve indicates a distance of about \(2.0~\text{\AA}\), identifying an inner-sphere species bound to a surface Co site. Only one O--H trace (red) stays at \(\approx1.0~\text{\AA}\), as expected for an OH group. The second-nearest H (blue) resides around \(1.6~\text{\AA}\) but sporadically dips to \(\approx1.0~\text{\AA}\), indicating transient protonation--deprotonation of the surface OH to form/re-form molecular water at the same site. These excursions are more frequent in the early part of the run, suggesting that the interfacial acid--base equilibrium is still relaxing before settling into a steady population of bound hydroxyls and outer-sphere water molecules.

\paragraph{Water migrating from the outer hydration layer to the epitaxial layer (B-termination).}
Figure~\ref{fig:PT}c shows a tagged water molecule that initially resides in the outer hydration region, with an O--Co distance of \(\sim4.5\!-\!5.0~\text{\AA}\) during the first \(\sim100\)~ps of the simulation. After this period, the molecule gradually moves closer to the surface and relocates into the epitaxial (outer-sphere) water layer, where its O--Co distance stabilizes around \(\sim3.3\!-\!3.6~\text{\AA}\). This relocation places the molecule in close interfacial proximity but still distinct from chemisorbed species, such as those in Figure~\ref{fig:PT}b, which bind at \(\sim2.0~\text{\AA}\). The H\(_1\)/H\(_2\) traces remain near \(\approx1.0~\text{\AA}\), confirming the molecule retains its molecular water character. Intermittent approaches of H\(_2\) toward \(\approx1.6~\text{\AA}\) indicate transiently strengthened hydrogen bonds with nearby OH/H\(_2\)O groups; these become somewhat more frequent after the relocation, suggesting enhanced interfacial exchange once the molecule has joined the epitaxial layer (see Figure~\ref{fig:Co-Ow_RDF}).

\paragraph{Water in the epitaxial layer with a late protonation event (B-termination).}
The tagged O--Co distance in Fig.~\ref{fig:PT}d is again \(\approx3.3\!-\!3.6~\text{\AA}\), diagnosing an outer-sphere water molecule at the interface. The most striking feature is the progressive approach of H\(_3\) (green) toward \(\approx1.0~\text{\AA}\) in the final tenth of the trajectory, while H\(_4\) (orange) simultaneously recedes, mirroring the bulk case in panel (a). This pattern is consistent with the formation of a hydronium-like configuration in the epitaxial layer near the end of the simulation and is corroborated by our coordination-count check, which flags frames where the tagged oxygen binds three hydrogens within a tight covalent cutoff.

Across (b--d), all examples are taken from the B-terminated interface for consistency. Where hydronium-like behavior is inferred, we verify that three O--H distances are simultaneously near \(1.0~\text{\AA}\) over multiple frames; isolated single-frame spikes can also arise from fluctuating hydrogen bonds and should not be over-interpreted. Likewise, an O--Co distance near \(2.0~\text{\AA}\) is a robust fingerprint of a chemisorbed OH/H\(_2\)O, whereas \(\approx3.3\!-\!3.6~\text{\AA}\) indicates outer-sphere binding in the epitaxial layer and is consistent with the second peak in the interfacial Co--O\(_\mathrm{w}\) RDF (Fig.~\ref{fig:RDF-2}a). Finally, the observation of late-time protonation events highlights the importance of long trajectories: such rare events are easily missed in short AIMD runs but emerge in extended classical MD sampling.

Overall, it can be concluded that (i) Bulk and interfacial protonation show the same mechanistic signature (three near-\(1.0~\text{\AA}\) O--H distances accompanied by the retreat of the fourth H), yet occur in different structural environments (no Co partner vs.\ outer-/inner-sphere binding). (ii) Surface OH groups undergo intermittent protonation/deprotonation while staying attached to Co, mediating charge/proton mobility at the interface. (iii) Waters entering the epitaxial layer can participate in enhanced hydrogen-bond rearrangements and, occasionally, hydronium-like episodes late in the trajectory, especially on the B-termination.

\subsection{Structure of the Contact Layers: A-termination vs. B-termination}\label{sec:Contact_Layer}

Due to the complexity and diverse chemistry at the contact layers, this section first presents several analyses, including atomic snapshots of the contact layers, trajectory lines of the surfaces and the first interfacial layers, and radial distribution functions (RDFs). The discussion then follows, based on the combined insights from these analyses.

\subsubsection{Snapshots, Trajectory Lines, and Radial Distribution Functions}\label{sec:Snapshots}

\begin{figure*}[ht]
\centering
\includegraphics[width=0.8\textwidth, trim=0 0 0 0, clip=true]{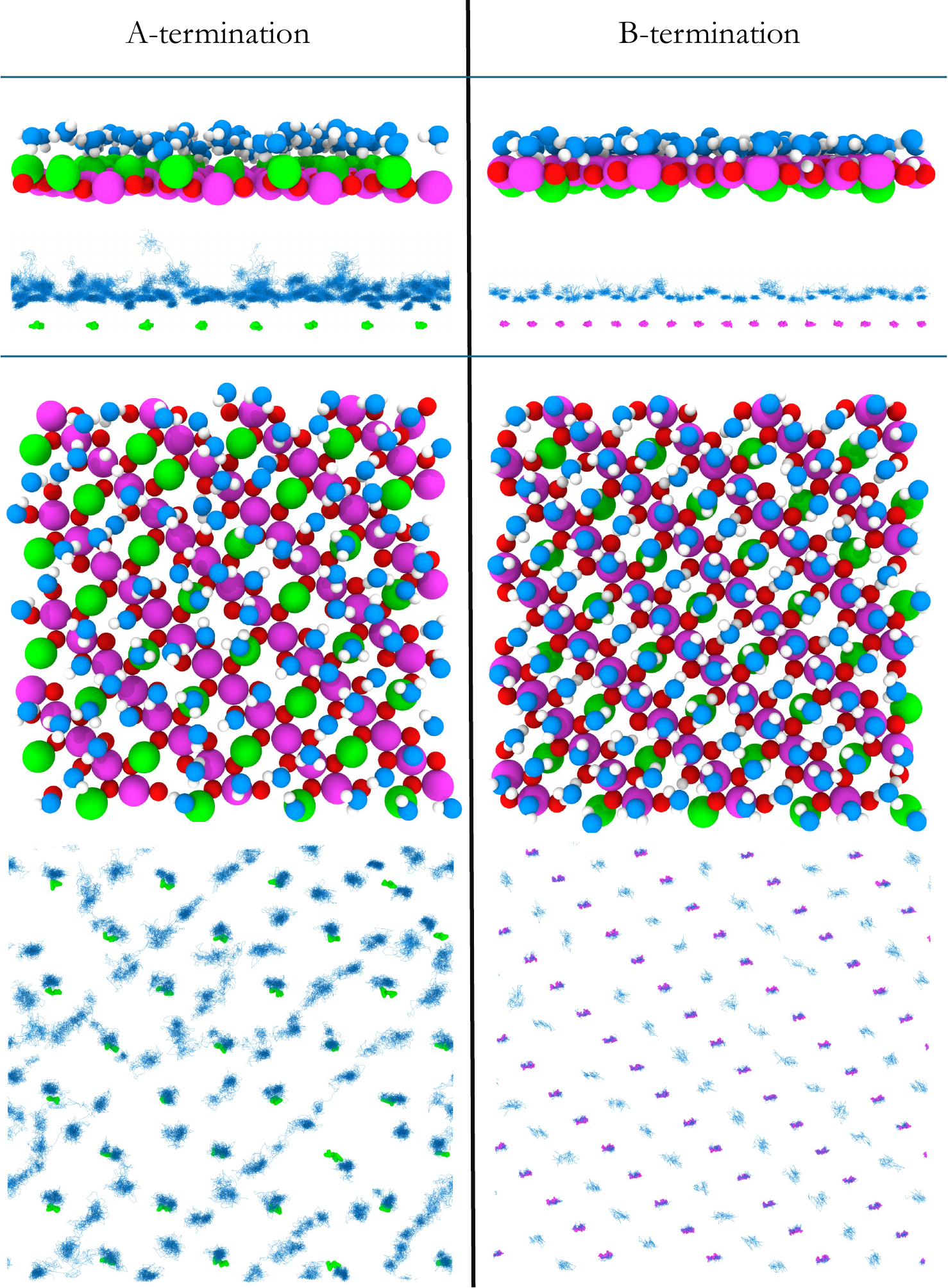}
\caption{
Contact layers of equilibrated Co$_3$O$_4$(001)--water interfaces at 300~K for the A-termination (left) and the B-termination (right).
Row~1: side views of the first interfacial layer and the top oxide layer.  
Row~2: trajectory overlays of surface Co atoms and interfacial oxygen species.  
Row~3: top views of the same region.  
Row~4: in-plane trajectories of first-layer surface Co atoms and interfacial oxygen species.  
Only a small interfacial portion of the slab is displayed for clarity.  
Color code: tetrahedral Co$^{2+}$ green, octahedral Co$^{3+}$ purple, lattice O red, interfacial oxygen species blue, H white.  
On the B-termination, water forms a compact, laterally ordered interfacial layer with well-defined adsorption sites, whereas the A-termination exhibits a more diffuse, weakly templated first layer.  
This means the A-termination does not strongly control where water molecules adsorb, leading to a disordered, loosely packed interfacial structure without clear registry to the surface lattice.  
First-layer Co$^{3+}$ trajectories are more localized than those of Co$^{2+}$, indicating stronger surface templating on the B-termination.}
\label{fig:snapshots}
\end{figure*}

Figure~\ref{fig:snapshots} illustrates the contact layers of the equilibrated Co$_3$O$_4$(001)--water interfaces at 300~K for the A-termination on the left and the B-termination on the right.
Row~1 shows side views of only the first interfacial layer with respect to the outermost surface plane. (Color code: tetrahedral Co$^{2+}$ green, octahedral Co$^{3+}$ purple, lattice O red, interfacial oxygen species blue, H white.)
The main difference between the A- and B-terminated surfaces is that, in the A-termination, the tetrahedral Co$^{2+}$ (green) are located at the surface and directly exposed to the interfacial layer, whereas in the B-termination, the octahedral Co$^{3+}$ (purple) occupy the surface and directly interact with the interfacial water layer.

Row~2 shows similar side views but overlays the trajectory lines of the surface Co atoms (Co$^{2+}$ green for the A-termination and Co$^{3+}$ purple for the B-termination) and the oxygen species of the first interfacial water layers over the last 50~ps of the simulation.  
Due to the highly dynamic and reactive nature of the solvent layer, it is difficult to clearly distinguish between H$_2$O molecules, OH$^-$ groups, and transient H$_3$O$^+$ species. The boundaries between these species are blurred near the interface; for example, a water molecule can exchange a proton with a surface oxygen, a neighboring OH$^-$, or another H$_2$O, and may even transiently share an excess proton.  
Therefore, in these snapshots, all oxygens of the interfacial layer are consistently shown in blue (in contrast to the red lattice oxygens). It should be kept in mind, however, that these blue oxygens may correspond to H$_2$O, OH$^-$, or H$_3$O$^+$. For a detailed mechanistic discussion, see Section~\ref{sec:Mechanism}.

Row~3 shows top views of the same first-layer region. Row~4 overlays the trajectories of the first-layer surface Co atoms and the interfacial oxygen species (similar to Row~3 but from the top view) to highlight their in-plane mobility.  
For visual clarity, only a small interfacial portion of the slab is shown in each panel. In addition, the trajectory lines of the interfacial oxygens are plotted thinner than the Co trajectories, so that they remain distinguishable.

To complement the above structural analysis and resolve local water reactivity and bonding, we analyze two key radial distribution functions (RDFs):  
(i) the RDF between surface cobalt atoms and oxygen species from any interfacial adsorbed species (Co$^{*}$–O$_\mathrm{w}$), where O$_\mathrm{w}$ includes oxygens from molecular water, hydroxyl groups, and transient hydronium ions; and  
(ii) the RDF between lattice surface oxygen atoms and hydrogen atoms (O$_\mathrm{s}$–H), which quantifies surface protonation via proton transfer from the interfacial layer.

It should be emphasized that, in the first RDF (Co$^{*}$–O$_\mathrm{w}$), the Co ions for the A-termination are Co$^{2+}$, while for the B-termination they are Co$^{3+}$.  
Thus, comparing the A- and B-terminations also reflects how different cobalt oxidation states influence the adsorption geometry and structuring of the interfacial layer.

\begin{figure}[H]
\centering
\begin{subfigure}{0.48\textwidth}
    \centering
    \caption{$\mathrm{Co}^{*}$–$\mathrm{O}_\mathrm{w}$}
    \label{fig:Co-Ow_RDF}
    \includegraphics[width=\textwidth, trim=0 0 0 0, clip=true]{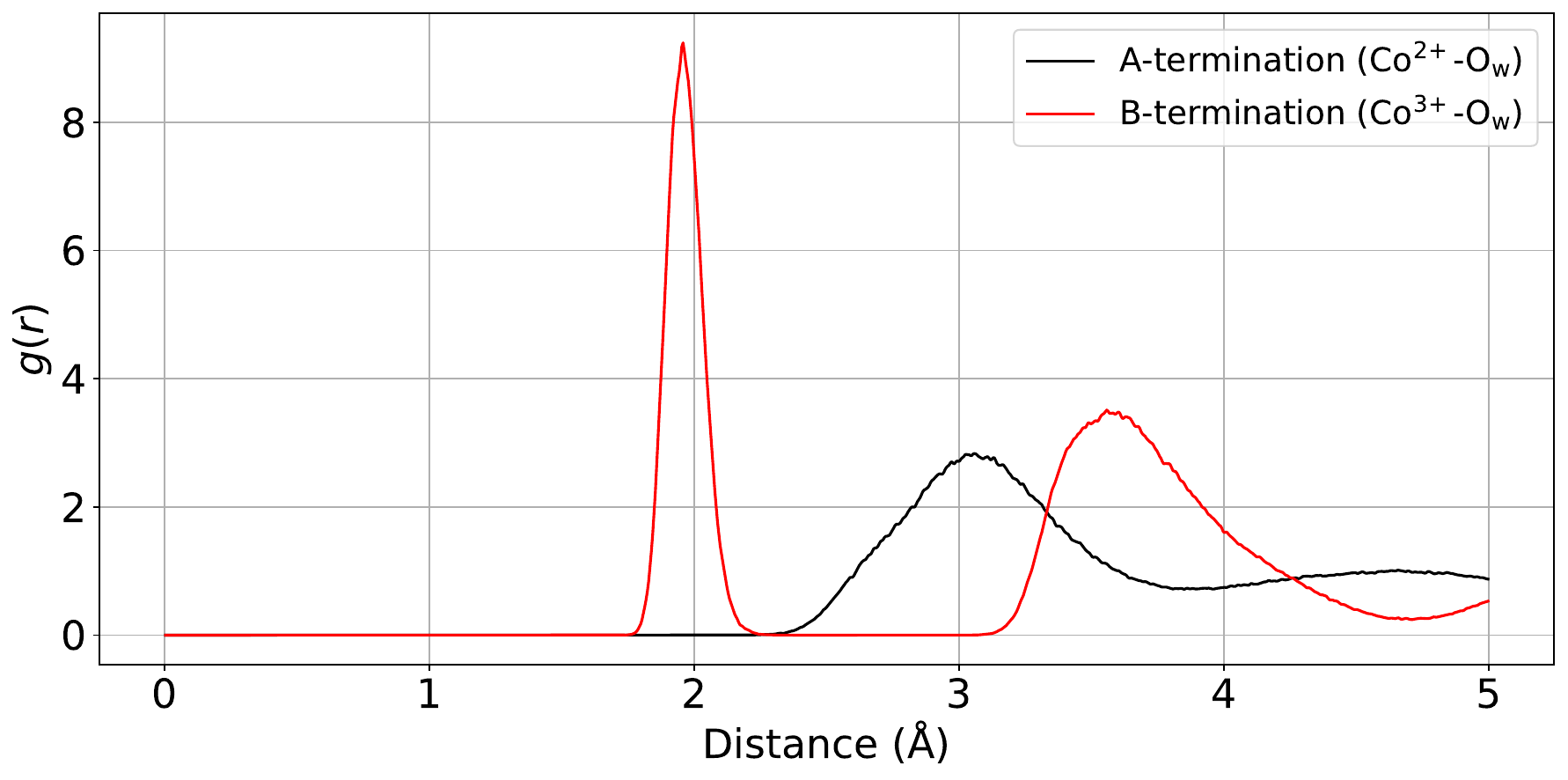}
\end{subfigure}%
\hfill
\begin{subfigure}{0.48\textwidth}
    \centering
    \caption{$\mathrm{O}_\mathrm{s}$–$\mathrm{H}$}
    \label{fig:Os-H_RDF}
    \includegraphics[width=\textwidth, trim=0 0 0 0, clip=true]{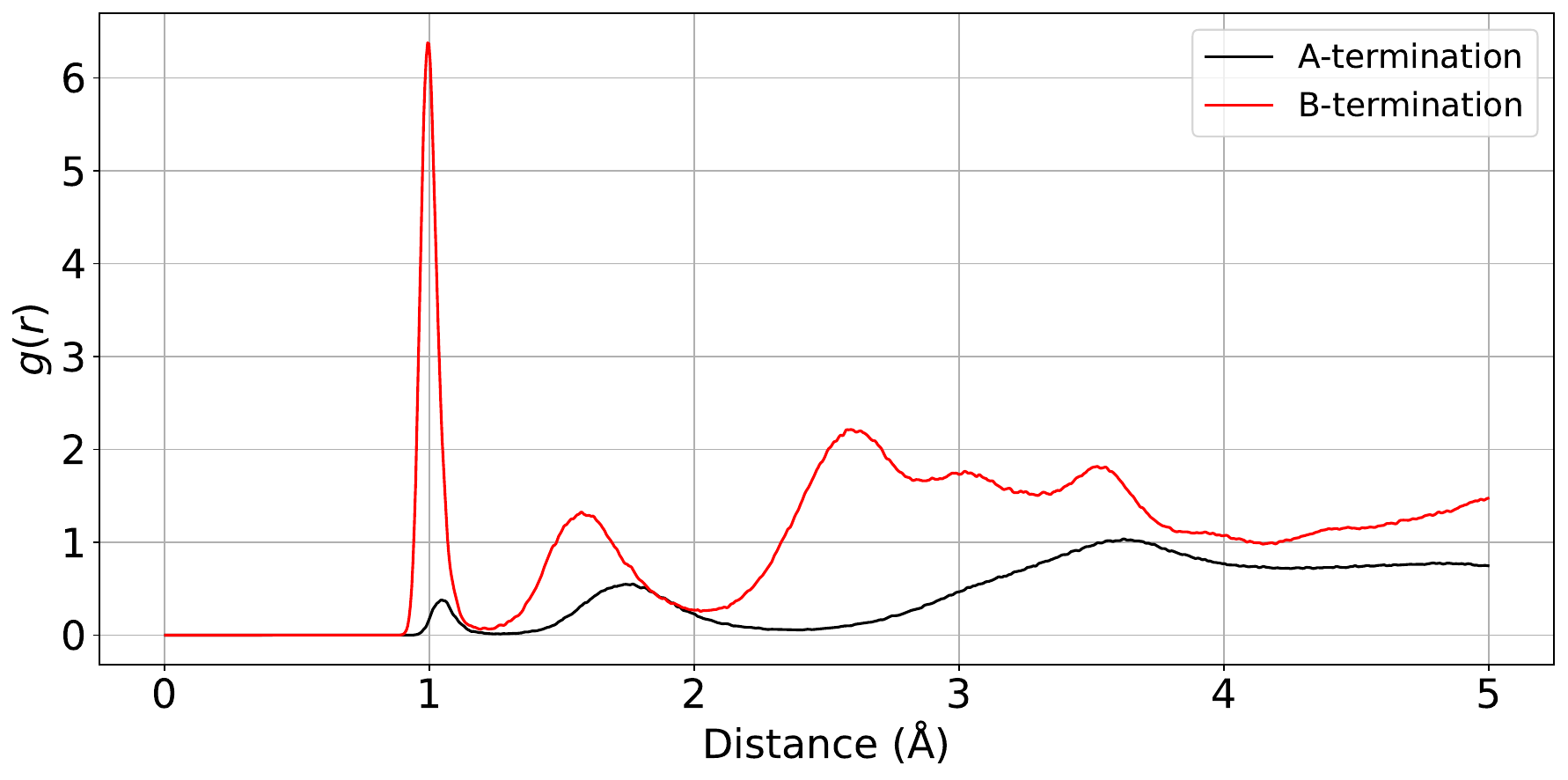}
\end{subfigure}%
\caption{
Radial distribution functions of the Co$_3$O$_4$(001) slab with A-termination (black) and B-termination (red).  
(a) Co$^{*}$–O$_\mathrm{w}$: the B-termination shows a sharp first maximum near 2~\AA{}, consistent with strong inner-sphere coordination, whereas the A-termination shows a broader, weaker feature shifted toward $\approx$~3~\AA{}.  
(b) O$_\mathrm{s}$–H: the B-termination exhibits a stronger $\approx$~1.0~\AA{} peak (more protonation) and a clearer hydrogen-bonding feature at 1.6–1.8~\AA{}.  
These RDFs confirm the stronger templating, higher hydroxylation, and more ordered interfacial layer already observed in the snapshots (Fig.~\ref{fig:snapshots}) and density profiles (Fig.~\ref{fig:density-profiles}).}
\label{fig:RDF-2}
\end{figure}

\textit{Co$^{*}$–O$_\mathrm{w}$ (Fig.~\ref{fig:RDF-2}a).}
On the B-termination, the Co$^{*}$–O$_\mathrm{w}$ RDF shows a sharp first maximum at \(r \approx 2.0~\text{\AA}\), followed by a deep minimum and a second maximum near \(3.3\!-\!3.6~\text{\AA}\).  
The A-termination instead exhibits a broader, weaker feature shifted to larger distances ($\approx$3~\AA{}), reflecting weaker and less specific adsorption of interfacial species.  
For reference, bulk Co–O distances for octahedral Co$^{3+}$ and tetrahedral Co$^{2+}$ are $\approx$1.95–1.96~\AA{}\cite{omranpour2024high}, bracketing the B-termination first maximum and confirming its inner-sphere character.  
Importantly, O$_\mathrm{w}$ here refers to all oxygen species in the interfacial layer (H$_2$O, OH$^-$, and H$_3$O$^+$), excluding lattice oxygens, i.e., not only molecular water alone.

\textit{O$_\mathrm{s}$–H (Fig.~\ref{fig:RDF-2}b).}
The O$_\mathrm{s}$–H RDF clearly separates covalently bound surface hydroxyls from hydrogen-bonded water.
A sharp peak near $\approx$1.0~\AA{} corresponds to surface OH groups formed via protonation by interfacial species, while a broader feature at 1.6–1.8~\AA{} originates from hydrogen bonding with water molecules in the interfacial layer.
The B-termination exhibits a much stronger OH peak and significantly more pronounced hydrogen bonding, demonstrating its higher reactivity toward water dissociation and the formation of a dense, oriented interfacial network.
In contrast, the A-termination shows considerably weaker protonation and a more disordered hydrogen-bond network, reflecting its substantially lower activity for water dissociation.

\clearpage

\subsubsection{Discussion}\label{sec:discussion_CL}

The first peak of the Co$^{*}$–O$_\mathrm{w}$ RDF (Fig.~\ref{fig:RDF-2}a), located at $\approx$2~\AA{}, coincides with the equilibrium Co–O distance in bulk Co$_3$O$_4$ and can thus be attributed to oxygens of chemisorbed species coordinated to surface Co$^{3+}$ ions on the B-termination.  

But what does the second peak represent?  
In Row~4 of Fig.~\ref{fig:snapshots}, two distinct mobilities of interfacial oxygen species (blue trajectories) can be identified:  
(i) oxygens located directly on top of surface Co atoms (purple) show reduced lateral motion, whereas  
(ii) oxygens positioned between these sites display significantly higher mobility.  
However, as visible in Rows~1 and~2 for the B-termination, both groups of oxygens belong to the same first interfacial layer at similar distances from the surface.  
The second peak can therefore be assigned to an outer-sphere (epitaxial) interfacial layer, consisting of oxygens laterally displaced relative to surface Co sites and hydrogen-bonded to chemisorbed OH groups or other interfacial species, rather than to a distinct, vertically separated second layer.

For the A-termination, the Co$^{*}$–O$_\mathrm{w}$ RDF (Fig.~\ref{fig:RDF-2}a) shows its first feature at a larger distance, $\approx$~3.0~\AA{}, indicating weaker binding of interfacial oxygen species to surface Co$^{2+}$ ions.  
This weaker interaction is also evident from the trajectory lines in Rows~2 and 4 of Fig.~\ref{fig:snapshots}, which show much higher mobility of the interfacial oxygen species  (both laterally and vertically) compared to the B-termination.  
Additionally, Row~3 of Fig.~\ref{fig:snapshots} shows a more disordered interfacial layer for the A-termination, with no clear adsorption registry relative to the underlying lattice.

In summary, a direct comparison reveals a pronounced difference between the two terminations:

\textit{B-termination}: the first interfacial water layer organizes into a compact, laterally ordered adlayer with well-defined adsorption sites exhibiting clear registry with the surface lattice.  
  In the top view, interfacial oxygen species cluster into recurring motifs with relatively uniform nearest-neighbor spacing, while in the side view the layer remains geometrically tight with reduced out-of-plane roughness.  
  
\textit{A-termination}: the first interfacial water layer is more diffuse and spatially heterogeneous.  
  Adsorption sites are more broadly distributed without a persistent repeating pattern, and the layer exhibits larger height fluctuations.  

These qualitative differences indicate a stronger templating effect and more specific binding geometries on the B-termination, whereas the A-termination exhibits a more liquid-like and weakly structured hydration environment. Moreover, the B-terminated surface also shows significantly higher degree of protonation (see Section~\ref{sec:Density Profiles}).

\subsection{Density Profiles}\label{sec:Density Profiles}

\begin{figure}[H]
  \centering
  \begin{subfigure}{0.48\textwidth}
    \centering
    \includegraphics[width=\textwidth,trim=0 0 0 0,clip]{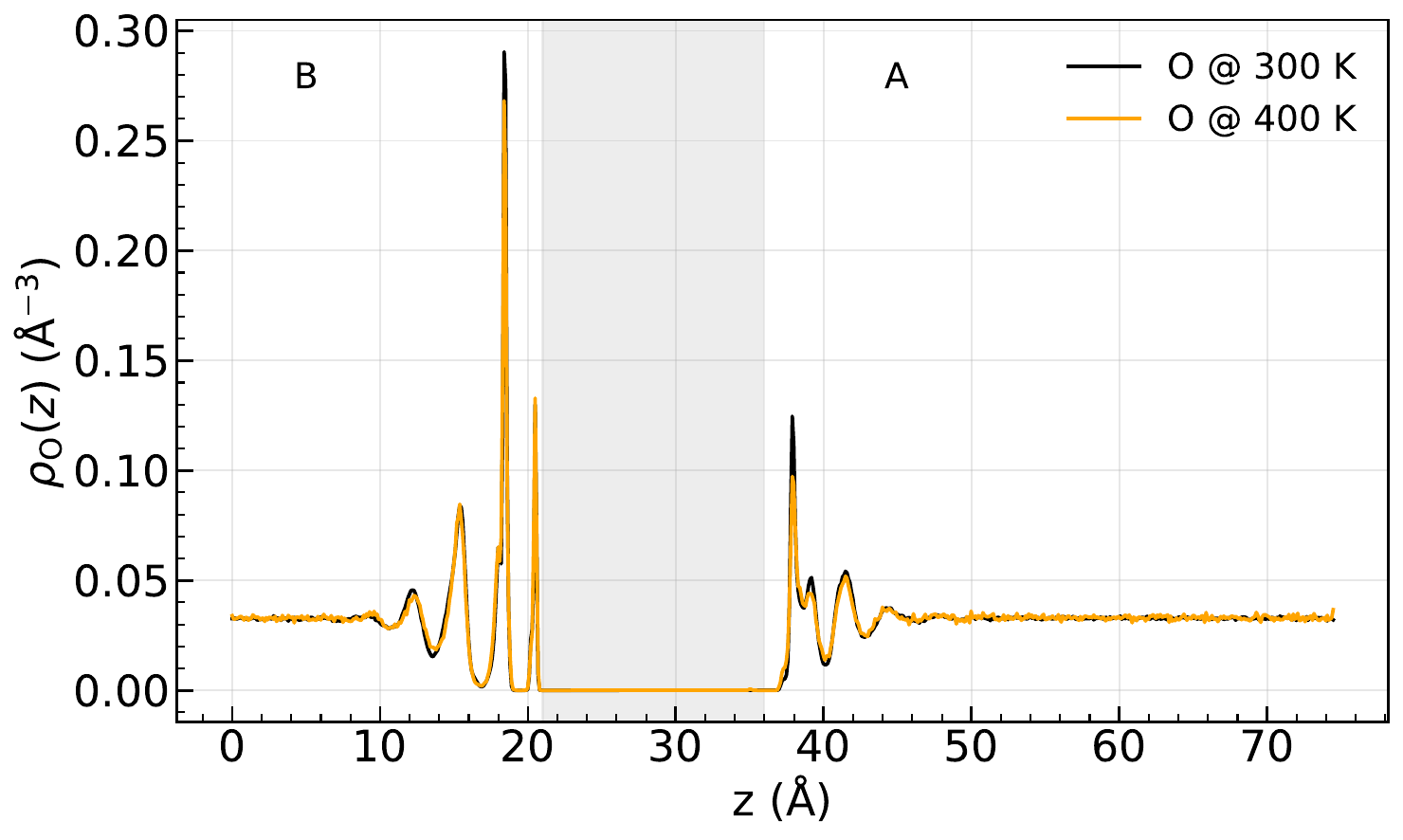}
    \caption{}
    \label{fig:O-density}
  \end{subfigure}%
  \hfill
  \begin{subfigure}{0.48\textwidth}
    \centering
    \includegraphics[width=\textwidth,trim=0 0 0 0,clip]{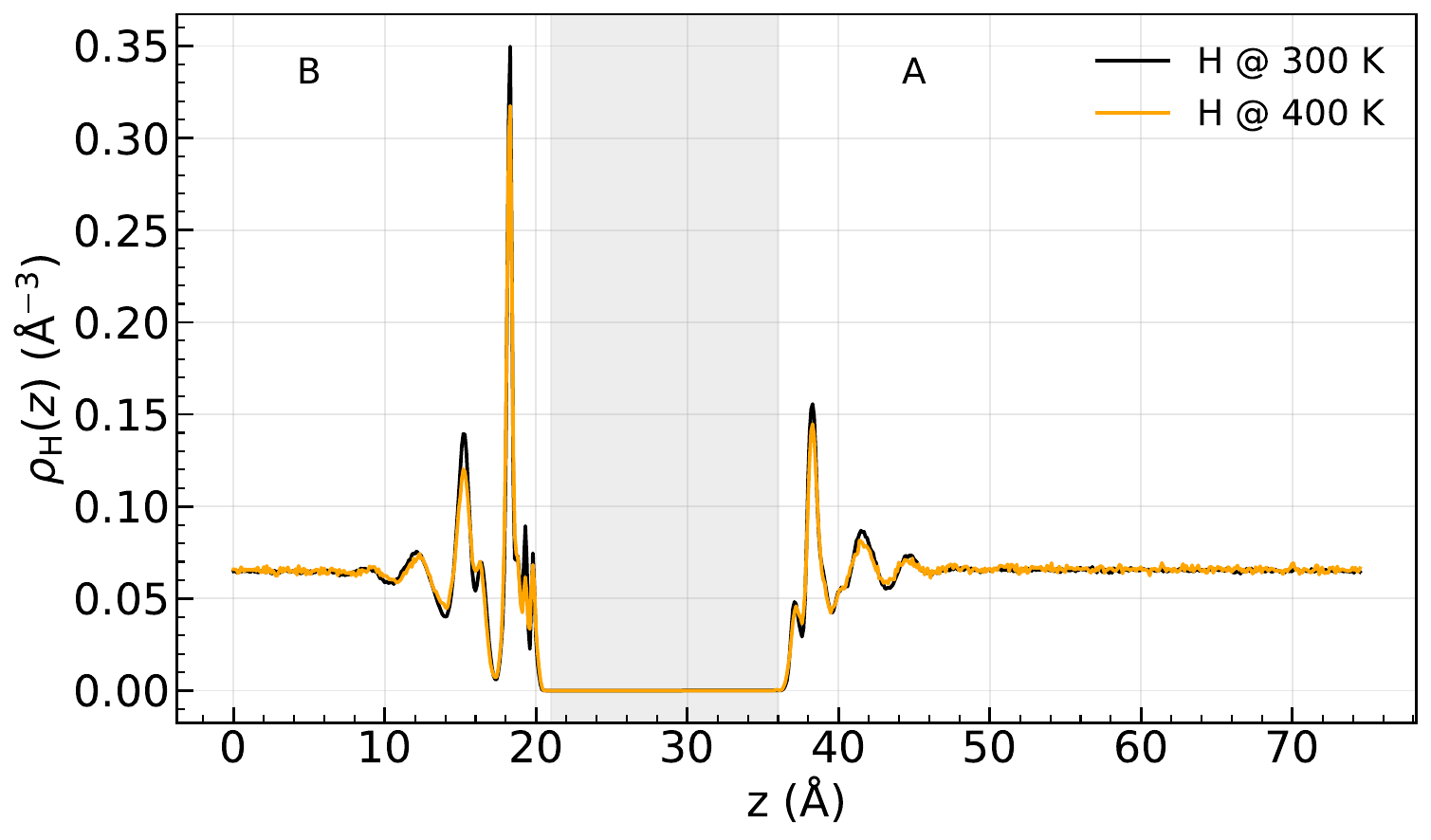}
    \caption{}
    \label{fig:H-density}
  \end{subfigure}
\caption{Water number density profiles along $z$ near the Co$_3$O$_4$(001) slab at $T=300$~K and $T=400$~K. 
Panel~(a) shows $\rho_{\mathrm{O}}(z)$ and panel~(b) shows $\rho_{\mathrm{H}}(z)$. 
The slab is exposed to water on both sides, resulting in two interfacial regions separated by a flat, water-free slab interior (gray region). 
At each interface, both profiles display a sharp contact-layer maximum followed by damped oscillations that relax to the bulk value beyond $\sim10$~\AA{}. 
The B-termination (left) structures interfacial water more strongly than the A-termination (right), as evidenced by a taller and narrower first maximum, a deeper first minimum, and a more pronounced second maximum. 
On the B side, the first O maximum partly arises from \emph{protonated surface oxygens}, which, by our definition, are counted as part of the ``water'' population. 
This explains why the O peak appears slightly closer to the slab than the first H peak. 
For molecular water, however, the H maximum still lies closer to the surface than the O maximum, consistent with preferential H-bond donation to surface oxygens. 
Increasing the temperature from 300~K to 400~K slightly lowers and broadens the peaks, consistent with enhanced thermal fluctuations. 
\emph{Selection rule:} an atom is counted if it has an O--H neighbor within $1.25$~\AA{}, which excludes lattice oxygens but includes protonated surface oxygens (OH groups).}
  \label{fig:density-profiles}
\end{figure}

The structural characteristics of interfacial water are analyzed by computing the number density profiles of oxygen and hydrogen atoms, as shown in Fig.~\ref{fig:density-profiles}. The density profiles $\rho_i(z)$ ($i=\mathrm{H},\mathrm{O}$) provide a spatially resolved view of how liquid water arranges itself at an oxide interface, capturing the contact layer, subsequent layering, and convergence to the bulk structure on the \AA{} scale. For Co$_3$O$_4$(001) slabs exposed to water on both faces, these profiles reveal how the two crystallographic terminations, A and B, template the adjacent liquid differently through the height, position, and sharpness of the first and second maxima, as well as how far this ordering propagates into the liquid phase. Comparing oxygen and hydrogen profiles also provides an orientational cue: H peaks that are closer to the surface than O peaks indicate water molecules donating H-bonds toward surface acceptors. It is important to note that according to the definition used in these plots, if a surface oxygen is protonated, it is also counted as an OH group.

\textit{Overall features and assignment.}  
Because the Co$_3$O$_4$(001) slab is exposed to water on both sides, the profiles in Fig.~\ref{fig:density-profiles} exhibit two interfacial regions separated by the slab interior (gray-shaded region). Reading from left to right, the first interfacial zone corresponds to the B-terminated surface and the second to the A-terminated surface. At each interface, both $\rho_{\mathrm{O}}(z)$ and $\rho_{\mathrm{H}}(z)$ display a sharp contact-layer maximum followed by damped oscillations that decay into a flat bulk plateau, characteristic of packing frustration and layering at solid–liquid interfaces.

\textit{A–termination vs. B–termination.}  
The main structural differences between the two terminations are revealed most clearly in the region of the second maxima, which dominate the interfacial water density. The B-termination exhibits a more pronounced and narrower second maximum, a deeper inter-layer minimum, and a slower decay toward the bulk plateau compared to the A-termination, reflecting stronger templating and tighter lateral ordering of the interfacial bilayer. The smaller first O peak visible at the B-side interface arises predominantly from protonated surface oxygens (surface OH groups), which are by definition counted as part of the ``water'' density. In contrast, the first peak on the A side is mostly molecular water, which explains why the apparent O peak height at the B interface is lower than at the A interface but does not contradict the stronger templating effect of the B-termination overall.

\textit{Hydrogen vs. oxygen: interfacial orientation and dissociation.}  
For molecular water at both terminations, the H maxima lie slightly closer to the surface than the corresponding O maxima, consistent with preferential donation of H-bonds toward surface oxygens. On the B-termination, however, an additional, smaller H peak appears even closer to the surface than the main molecular H maximum. This peak originates from \emph{dissociated protons bound to surface oxygens} (forming OH groups), as confirmed by the strong Os–H RDF peak at $\sim1.0$~\AA{} (Fig.~\ref{fig:RDF-2}). Thus, for the B-termination, the innermost hydrogen density partly reflects surface protonation rather than orientational ordering of intact water molecules. This resolves the apparent mismatch between O and H peak positions: the smaller first O peak is dominated by protonated surface species, while the main second O/H peaks correspond to intact molecular water.

\textit{Bulk region and normalization.}  
Far from either surface, both $\rho_{\mathrm{O}}(z)$ and $\rho_{\mathrm{H}}(z)$ approach constant plateaus with the expected stoichiometric ratio $\rho_{\mathrm{H}}^{\mathrm{bulk}}\!\approx\!2\,\rho_{\mathrm{O}}^{\mathrm{bulk}}$ of bulk water. The flatness of these plateaus and the correct 2:1 ratio confirm proper normalization, adequate sampling, and negligible finite-size artifacts.

\textit{Temperature dependence (300\,K vs.\ 400\,K).}  
Raising the temperature from $300$~K to $400$~K reduces the height and sharpness of the interfacial peaks, consistent with enhanced thermal fluctuations and increased mobility.  
Before the production runs, the systems were equilibrated in the NPT ensemble at their respective temperatures, so the simulation cell volumes are slightly different at 300~K and 400~K.  
Consequently, small differences in the bulk density plateaus are expected and consistent with the setup described in Section~\ref{sec:Computational}.  
Importantly, the qualitative ordering persists: the B-termination maintains stronger interfacial structuring than the A-termination at both temperatures.

In summary:
\begin{enumerate}
  \item Water at Co$_3$O$_4$(001) forms a structured interfacial bilayer at both terminations, followed by damped oscillations that converge to bulk density within a few~\AA.
  \item The B-terminated surface supports a denser, more laterally ordered interfacial structure than the A-termination, as seen from its sharper secondary maxima, deeper minima, and slower decay.
  \item The innermost O and H peaks on the B side originate mainly from protonated surface oxygens (OH groups), while the main O/H peaks farther from the surface correspond to intact molecular water donating H-bonds. Thus, the additional close-in H peak on the B side reflects surface hydroxylation rather than molecular orientation.
  \item Increasing the temperature slightly reduces interfacial ordering but preserves bulk stoichiometry and the strong B\,>\,A contrast in templating.
\end{enumerate}

Together with the discussion of the previous section, these density profiles confirm a consistent picture:  
the B-terminated Co$_3$O$_4$(001) surface forms a more ordered, tightly bound interfacial bilayer of partially dissociated water molecules that both hydroxylates and hydrogen-bonds to the surface,  
while the A-termination supports a looser and more weakly structured hydration layer.  
This ordered, quasi-epitaxial water adlayer on the B-termination has been proposed to enhance oxidative catalytic activity~\cite{P6444,kox2020impact}.

\subsection{Discussion on the Implications for Catalysis}\label{sec:implications}

As mentioned in the Introduction, the broader context of this work lies in catalysis in the aqueous phase in contact with solid surfaces and aims to elucidate a deeper understanding of such complex interfaces. Performing reactions such as alcohol oxidation in water can improve catalytic selectivity and enable catalyst reusability, both of which are central to practical catalysis. The interaction between water and transition-metal oxides (TMOs) such as Co$_3$O$_4$ is, however, more complex than at metal/water interfaces~\cite{henderson2002interaction,björneholm2016water}. Water plays a multifaceted role, with both beneficial and detrimental consequences for oxidation chemistry~\cite{davies2016role,lin2021heterogeneous}. On the beneficial side, it can facilitate proton-coupled steps by providing mobile proton carriers (transient OH$^-$, H$_3$O$^+$, and surface hydroxyls) that temporarily store and relay protons~\cite{P6444,stuchebrukhov2010theory,agmon2016protons}; it has also been shown to promote O$_2$ activation in certain oxidation regimes~\cite{munoz2018solvation,tran2016water}. On the detrimental side, water competes with reactants (e.g., alcohols) for adsorption sites and can also passivate lattice oxygen functionality~\cite{P7141}, thereby diminishing surface reactivity. Which of these tendencies dominates is highly dependent on the reaction and conditions (pH, potential, temperature, coverage, defects, termination) to be investigated in future investigations.

This work specifically sought to bridge the gap between previous \textit{ab initio} simulations and experimental observations and contributes to the understanding of Co$_3$O$_4$(001)--water interfaces.  
It confirmed several earlier findings while elucidating additional features of the Co$_3$O$_4$(001) A- and B-terminations, in particular the higher activity of the B-termination for water dissociation.  
This conclusion is directly supported by the O$_s$--H RDF (Sec.~\ref{sec:Contact_Layer}), which shows a significantly stronger $\approx$1.0~\AA{} peak on the B side, indicating more extensive protonation of surface oxygens compared to the A-termination.  
Furthermore, the Co$^\ast$--O$_\mathrm{w}$ RDF clearly reveals how the oxidation state of the surface cobalt ions affects their interaction with water: surface Co$^{3+}$ ions on the B-termination exhibit stronger inner-sphere coordination with interfacial oxygens (sharp $\approx$2.0~\AA{} peak), whereas surface Co$^{2+}$ ions on the A-termination bind more weakly, resulting in a broader, more distant $\approx$3.0~\AA{} feature.  
However, it should be emphasized that higher interfacial water structuring or enhanced hydroxylation does not necessarily translate into superior catalytic performance for all reactions; for example, in alcohol oxidation, many competing factors are at play.  
As our analysis showed (Sec.~\ref{sec:Mechanism}), multiple microscopic pathways operate simultaneously near the interface (e.g., proton transfer between waters, protonation/deprotonation of chemisorbed hydroxyls, and exchange between inner- and outer-sphere species), and introducing specific reactants only adds further complexity.  
These results reinforce the need for a more complete atomistic-level description of the Co$_3$O$_4$--water interface as a prerequisite for predictive understanding and optimization of aqueous-phase oxidation catalysis.

\section{Conclusions}\label{sec:conclusions}

In this work, we developed and applied a high-dimensional neural network potential to investigate the structure, dynamics, and reactivity of Co$_3$O$_4$(001)--water interfaces. By combining density functional theory reference data with active learning, we constructed a robust potential that enables nanosecond-scale molecular dynamics simulations for systems comprising thousands of atoms, thereby overcoming the spatial and temporal limitations of previous \textit{ab initio} studies.

Our simulations reveal pronounced differences between the two possible surface terminations. The B-terminated Co$_3$O$_4$(001) surface forms a highly ordered, quasi-epitaxial hydration layer with strong templating effects, enhanced hydroxylation, and well-defined hydrogen-bonding networks. In contrast, the A-termination exhibits a more diffuse contact layer with weaker templating, lower hydroxylation, and less structured interfacial water. These structural differences are directly linked to the distinct reactivity of the terminations: the B-terminated surface shows a higher propensity for water dissociation, consistent with experimental observations of its greater catalytic activity.

Furthermore, the extended timescales captured in our simulations allow us to identify rare events that are inaccessible to shorter \textit{ab initio} molecular dynamics runs. These include late-time proton transfer episodes within the outer hydration layer and dynamic protonation–deprotonation of surface hydroxyl groups, which play an essential role in interfacial charge and proton transport. Temperature-dependent simulations demonstrate that increasing the temperature reduces interfacial ordering but preserves the stronger templating and enhanced reactivity of the B-terminated surface.

Overall, this study provides an atomistically detailed and dynamically resolved picture of Co$_3$O$_4$(001)--water interfaces under realistic solvation conditions. The insights gained into interfacial structure, hydroxylation, and proton transfer mechanisms bridge the gap between experimental findings (particularly regarding the crucial role of Co$^{3+}$)~\cite{P6391} and previous \textit{ab initio} molecular dynamics simulations~\cite{kox2020impact}.

\section*{Acknowledgments}
We are grateful for funding by the Deutsche Forschungsgemeinschaft (DFG, German Research Foundation) in TRR/CRC 247 (A10, project-ID 388390466) and under Germany's Excellence Strategy – EXC 2033 RESOLV (project-ID 390677874). The authors also acknowledge the computing time provided to them by the Paderborn Center for Parallel Computing (PC2).

\section*{Data Availability}
The data generated and analyzed during the current study are available from the corresponding authors upon request.
\bibliography{literature}

\providecommand{\latin}[1]{#1}
\makeatletter
\providecommand{\doi}
  {\begingroup\let\do\@makeother\dospecials
  \catcode`\{=1 \catcode`\}=2 \doi@aux}
\providecommand{\doi@aux}[1]{\endgroup\texttt{#1}}
\makeatother
\providecommand*\mcitethebibliography{\thebibliography}
\csname @ifundefined\endcsname{endmcitethebibliography}
  {\let\endmcitethebibliography\endthebibliography}{}
\begin{mcitethebibliography}{93}
\providecommand*\natexlab[1]{#1}
\providecommand*\mciteSetBstSublistMode[1]{}
\providecommand*\mciteSetBstMaxWidthForm[2]{}
\providecommand*\mciteBstWouldAddEndPuncttrue
  {\def\EndOfBibitem{\unskip.}}
\providecommand*\mciteBstWouldAddEndPunctfalse
  {\let\EndOfBibitem\relax}
\providecommand*\mciteSetBstMidEndSepPunct[3]{}
\providecommand*\mciteSetBstSublistLabelBeginEnd[3]{}
\providecommand*\EndOfBibitem{}
\mciteSetBstSublistMode{f}
\mciteSetBstMaxWidthForm{subitem}{(\alph{mcitesubitemcount})}
\mciteSetBstSublistLabelBeginEnd
  {\mcitemaxwidthsubitemform\space}
  {\relax}
  {\relax}

\bibitem[Anke \latin{et~al.}(2019)Anke, Bendt, Sinev, Hajiyani, Antoni,
  Zegkinoglou, Jeon, Pentcheva, Roldan~Cuenya, Schulz, \latin{et~al.}
  others]{P6389}
Anke,~S.; Bendt,~G.; Sinev,~I.; Hajiyani,~H.; Antoni,~H.; Zegkinoglou,~I.;
  Jeon,~H.; Pentcheva,~R.; Roldan~Cuenya,~B.; Schulz,~S.; others Selective
  2-propanol oxidation over unsupported \ce{ \ce{Co3O4}} spinel nanoparticles:
  mechanistic insights into aerobic oxidation of alcohols. \emph{ACS Catal.}
  \textbf{2019}, \emph{9}, 5974--5985\relax
\mciteBstWouldAddEndPuncttrue
\mciteSetBstMidEndSepPunct{\mcitedefaultmidpunct}
{\mcitedefaultendpunct}{\mcitedefaultseppunct}\relax
\EndOfBibitem
\bibitem[Doheim and El-Shobaky(2002)Doheim, and El-Shobaky]{P7139}
Doheim,~M.; El-Shobaky,~H. Catalytic conversion of ethanol and iso-propanol
  over \ce{ZnO}-treated \ce{ \ce{Co3O4}}/\ce{Al2O3} solids. \emph{Colloids
  Surf.} \textbf{2002}, \emph{204}, 169--174\relax
\mciteBstWouldAddEndPuncttrue
\mciteSetBstMidEndSepPunct{\mcitedefaultmidpunct}
{\mcitedefaultendpunct}{\mcitedefaultseppunct}\relax
\EndOfBibitem
\bibitem[Yang \latin{et~al.}(2021)Yang, Kastenmeier, Ronovsk{\`y}, Fusek,
  Sk{\'a}la, Waidhas, Bertram, Tsud, Matvija, Prince, \latin{et~al.}
  others]{P7140}
Yang,~T.; Kastenmeier,~M.; Ronovsk{\`y},~M.; Fusek,~L.; Sk{\'a}la,~T.;
  Waidhas,~F.; Bertram,~M.; Tsud,~N.; Matvija,~P.; Prince,~K.~C.; others
  Selective electrooxidation of 2-propanol on Pt nanoparticles supported on
  \ce{ \ce{Co3O4}}: an in-situ study on atomically defined model systems.
  \emph{J. Phys. D} \textbf{2021}, \emph{54}, 164002\relax
\mciteBstWouldAddEndPuncttrue
\mciteSetBstMidEndSepPunct{\mcitedefaultmidpunct}
{\mcitedefaultendpunct}{\mcitedefaultseppunct}\relax
\EndOfBibitem
\bibitem[Omranpoor \latin{et~al.}(2023)Omranpoor, Bera, Bullert, Linke,
  Salamon, Webers, Wende, Hasselbrink, Spohr, and Kenmoe]{P7141}
Omranpoor,~A.~H.; Bera,~A.; Bullert,~D.; Linke,~M.; Salamon,~S.; Webers,~S.;
  Wende,~H.; Hasselbrink,~E.; Spohr,~E.; Kenmoe,~S. 2-Propanol interacting with
  \ce{Co3O4} (001): A combined vSFS and AIMD study. \emph{J. Chem. Phys.}
  \textbf{2023}, \emph{158}\relax
\mciteBstWouldAddEndPuncttrue
\mciteSetBstMidEndSepPunct{\mcitedefaultmidpunct}
{\mcitedefaultendpunct}{\mcitedefaultseppunct}\relax
\EndOfBibitem
\bibitem[Anke \latin{et~al.}(2020)Anke, Falk, Bendt, Sinev, Haevecker, Antoni,
  Zegkinoglou, Jeon, Knop-Gericke, Schl{\"o}gl, \latin{et~al.} others]{P6435}
Anke,~S.; Falk,~T.; Bendt,~G.; Sinev,~I.; Haevecker,~M.; Antoni,~H.;
  Zegkinoglou,~I.; Jeon,~H.; Knop-Gericke,~A.; Schl{\"o}gl,~R.; others On the
  reversible deactivation of cobalt ferrite spinel nanoparticles applied in
  selective 2-propanol oxidation. \emph{J. Catal.} \textbf{2020}, \emph{382},
  57--68\relax
\mciteBstWouldAddEndPuncttrue
\mciteSetBstMidEndSepPunct{\mcitedefaultmidpunct}
{\mcitedefaultendpunct}{\mcitedefaultseppunct}\relax
\EndOfBibitem
\bibitem[Douma \latin{et~al.}(2023)Douma, Nono, Omranpoor, Lamperti,
  Debernardi, and Kenmoe]{P6405}
Douma,~D.~H.; Nono,~K.~N.; Omranpoor,~A.~H.; Lamperti,~A.; Debernardi,~A.;
  Kenmoe,~S. Probing the local environment of active sites during 2-propanol
  oxidation to acetone on the \ce{Co3O4} (001) surface: Insights from
  first-principles O K-edge XANES spectroscopy. \emph{J. Phys. Chem. C}
  \textbf{2023}, \emph{127}, 5351--5357\relax
\mciteBstWouldAddEndPuncttrue
\mciteSetBstMidEndSepPunct{\mcitedefaultmidpunct}
{\mcitedefaultendpunct}{\mcitedefaultseppunct}\relax
\EndOfBibitem
\bibitem[Omranpoor \latin{et~al.}(2022)Omranpoor, Kox, Spohr, and
  Kenmoe]{P6444}
Omranpoor,~A.; Kox,~T.; Spohr,~E.; Kenmoe,~S. Influence of temperature, surface
  composition and electrochemical environment on 2-propanol decomposition at
  the \ce{Co3O4} (001)/\ce{H2O} interface. \emph{Appl. Surf. Sci. Advances}
  \textbf{2022}, \emph{12}, 100319\relax
\mciteBstWouldAddEndPuncttrue
\mciteSetBstMidEndSepPunct{\mcitedefaultmidpunct}
{\mcitedefaultendpunct}{\mcitedefaultseppunct}\relax
\EndOfBibitem
\bibitem[Kenmoe \latin{et~al.}(2022)Kenmoe, Douma, Raji, M'Passi-Mabiala,
  Götsch, Girgsdies, Knop-Gericke, Schlögl, and Spohr]{P6387}
Kenmoe,~S.; Douma,~D.~H.; Raji,~A.~T.; M'Passi-Mabiala,~B.; Götsch,~T.;
  Girgsdies,~F.; Knop-Gericke,~A.; Schlögl,~R.; Spohr,~E. {X-ray Absorption
  Near-Edge Structure (XANES) at the O K-Edge of Bulk \ce{ \ce{Co3O4}}:
  Experimental and Theoretical Studies}. \emph{Nanomaterials} \textbf{2022},
  \emph{12}, 921\relax
\mciteBstWouldAddEndPuncttrue
\mciteSetBstMidEndSepPunct{\mcitedefaultmidpunct}
{\mcitedefaultendpunct}{\mcitedefaultseppunct}\relax
\EndOfBibitem
\bibitem[Omranpoor and Kenmoe(2023)Omranpoor, and Kenmoe]{P7142}
Omranpoor,~A.~H.; Kenmoe,~S. 2-Propanol Activation on the Low Index \ce{Co3O4}
  Surfaces: A Comparative Study Using Molecular Dynamics Simulations.
  \emph{Catalysts} \textbf{2023}, \emph{14}, 25\relax
\mciteBstWouldAddEndPuncttrue
\mciteSetBstMidEndSepPunct{\mcitedefaultmidpunct}
{\mcitedefaultendpunct}{\mcitedefaultseppunct}\relax
\EndOfBibitem
\bibitem[Falk \latin{et~al.}(2021)Falk, Budiyanto, Dreyer, Pflieger, Waffel,
  B{\"u}ker, Weidenthaler, Ortega, Behrens, T{\"u}ys{\"u}z, \latin{et~al.}
  others]{P6391}
Falk,~T.; Budiyanto,~E.; Dreyer,~M.; Pflieger,~C.; Waffel,~D.; B{\"u}ker,~J.;
  Weidenthaler,~C.; Ortega,~K.~F.; Behrens,~M.; T{\"u}ys{\"u}z,~H.; others
  Identification of active sites in the catalytic oxidation of 2-propanol over
  Co1+ xFe2--xO4 spinel oxides at solid/liquid and solid/gas interfaces.
  \emph{ChemCatChem} \textbf{2021}, \emph{13}, 2942--2951\relax
\mciteBstWouldAddEndPuncttrue
\mciteSetBstMidEndSepPunct{\mcitedefaultmidpunct}
{\mcitedefaultendpunct}{\mcitedefaultseppunct}\relax
\EndOfBibitem
\bibitem[Feizi \latin{et~al.}(2019)Feizi, Bagheri, Song, Shen, Allakhverdiev,
  and Najafpour]{feizi2019cobalt}
Feizi,~H.; Bagheri,~R.; Song,~Z.; Shen,~J.-R.; Allakhverdiev,~S.~I.;
  Najafpour,~M.~M. Cobalt/cobalt oxide surface for water oxidation. \emph{ACS
  Sustain. Chem. Eng.} \textbf{2019}, \emph{7}, 6093--6105\relax
\mciteBstWouldAddEndPuncttrue
\mciteSetBstMidEndSepPunct{\mcitedefaultmidpunct}
{\mcitedefaultendpunct}{\mcitedefaultseppunct}\relax
\EndOfBibitem
\bibitem[Jiao and Frei(2009)Jiao, and Frei]{jiao2009nanostructured}
Jiao,~F.; Frei,~H. Nanostructured cobalt oxide clusters in mesoporous silica as
  efficient oxygen-evolving catalysts. \emph{Angew. Chem.} \textbf{2009},
  \emph{121}, 1873--1876\relax
\mciteBstWouldAddEndPuncttrue
\mciteSetBstMidEndSepPunct{\mcitedefaultmidpunct}
{\mcitedefaultendpunct}{\mcitedefaultseppunct}\relax
\EndOfBibitem
\bibitem[Hu \latin{et~al.}(2008)Hu, Peng, and Li]{hu2008selective}
Hu,~L.; Peng,~Q.; Li,~Y. Selective synthesis of \ce{Co3O4} nanocrystal with
  different shape and crystal plane effect on catalytic property for methane
  combustion. \emph{J. Am. Chem. Soc.} \textbf{2008}, \emph{130},
  16136--16137\relax
\mciteBstWouldAddEndPuncttrue
\mciteSetBstMidEndSepPunct{\mcitedefaultmidpunct}
{\mcitedefaultendpunct}{\mcitedefaultseppunct}\relax
\EndOfBibitem
\bibitem[Xie \latin{et~al.}(2009)Xie, Li, Liu, Haruta, and Shen]{xie2009low}
Xie,~X.; Li,~Y.; Liu,~Z.-Q.; Haruta,~M.; Shen,~W. Low-temperature oxidation of
  {CO} catalysed by \ce{Co3O4} nanorods. \emph{Nature} \textbf{2009},
  \emph{458}, 746--749\relax
\mciteBstWouldAddEndPuncttrue
\mciteSetBstMidEndSepPunct{\mcitedefaultmidpunct}
{\mcitedefaultendpunct}{\mcitedefaultseppunct}\relax
\EndOfBibitem
\bibitem[Li \latin{et~al.}(2005)Li, Xu, and Chen]{li2005co3o4}
Li,~W.-Y.; Xu,~L.-N.; Chen,~J. \ce{Co3O4} nanomaterials in lithium-ion
  batteries and gas sensors. \emph{Adv. Funct. Mater.} \textbf{2005},
  \emph{15}, 851--857\relax
\mciteBstWouldAddEndPuncttrue
\mciteSetBstMidEndSepPunct{\mcitedefaultmidpunct}
{\mcitedefaultendpunct}{\mcitedefaultseppunct}\relax
\EndOfBibitem
\bibitem[Waidhas \latin{et~al.}(2020)Waidhas, Haschke, Khanipour, Fromm,
  Görling, Bachmann, Katsounaros, Mayrhofer, Brummel, and
  Libuda]{waidhas2020secondary}
Waidhas,~F.; Haschke,~S.; Khanipour,~P.; Fromm,~L.; Görling,~A.; Bachmann,~J.;
  Katsounaros,~I.; Mayrhofer,~K.~J.; Brummel,~O.; Libuda,~J. Secondary Alcohols
  as Rechargeable Electrofuels: Electrooxidation of Isopropyl Alcohol at Pt
  Electrodes. \emph{ACS Catal.} \textbf{2020}, \emph{10}, 6831--6842\relax
\mciteBstWouldAddEndPuncttrue
\mciteSetBstMidEndSepPunct{\mcitedefaultmidpunct}
{\mcitedefaultendpunct}{\mcitedefaultseppunct}\relax
\EndOfBibitem
\bibitem[Hill and Hartwig(2017)Hill, and Hartwig]{hill2017site}
Hill,~C.~K.; Hartwig,~J.~F. Site-selective oxidation, amination and
  epimerization reactions of complex polyols enabled by transfer hydrogenation.
  \emph{Nat. Chem.} \textbf{2017}, \emph{9}, 1213--1221\relax
\mciteBstWouldAddEndPuncttrue
\mciteSetBstMidEndSepPunct{\mcitedefaultmidpunct}
{\mcitedefaultendpunct}{\mcitedefaultseppunct}\relax
\EndOfBibitem
\bibitem[Finocchio \latin{et~al.}(1997)Finocchio, Willey, Busca, and
  Lorenzelli]{finocchio1997ftir}
Finocchio,~E.; Willey,~R.~J.; Busca,~G.; Lorenzelli,~V. {FTIR} studies on the
  selective oxidation and combustion of light hydrocarbons at metal oxide
  surfaces Part 3.--Comparison of the oxidation of {C}\textsubscript{3} organic
  compounds over \ce{ \ce{Co3O4}}, \ce{MgCr\textsubscript{2}O\textsubscript{4}}
  and \ce{CuO}. \emph{J. Chem. Soc., Faraday Trans.} \textbf{1997}, \emph{93},
  175--180\relax
\mciteBstWouldAddEndPuncttrue
\mciteSetBstMidEndSepPunct{\mcitedefaultmidpunct}
{\mcitedefaultendpunct}{\mcitedefaultseppunct}\relax
\EndOfBibitem
\bibitem[Wang \latin{et~al.}(2012)Wang, Kavanagh, Guo, Guo, Lu, and
  Hu]{wang2012structural}
Wang,~H.-F.; Kavanagh,~R.; Guo,~Y.-L.; Guo,~Y.; Lu,~G.-Z.; Hu,~P. Structural
  origin: water deactivates metal oxides to CO oxidation and promotes
  low-temperature CO oxidation with metals. \emph{Angew. Chem. Int. Ed.}
  \textbf{2012}, \emph{51}, 6657--6661\relax
\mciteBstWouldAddEndPuncttrue
\mciteSetBstMidEndSepPunct{\mcitedefaultmidpunct}
{\mcitedefaultendpunct}{\mcitedefaultseppunct}\relax
\EndOfBibitem
\bibitem[Zasada \latin{et~al.}(2011)Zasada, Piskorz, Stelmachowski, Kotarba,
  Paul, Plocinnski, Kurzydlowski, and Sojka]{zasada2011periodic}
Zasada,~F.; Piskorz,~W.; Stelmachowski,~P.; Kotarba,~A.; Paul,~J.-F.;
  Plocinnski,~T.; Kurzydlowski,~K.~J.; Sojka,~Z. Periodic DFT and HR-STEM
  studies of surface structure and morphology of cobalt spinel nanocrystals.
  Retrieving 3D shapes from 2D images. \emph{J. Phys. Chem. C} \textbf{2011},
  \emph{115}, 6423--6432\relax
\mciteBstWouldAddEndPuncttrue
\mciteSetBstMidEndSepPunct{\mcitedefaultmidpunct}
{\mcitedefaultendpunct}{\mcitedefaultseppunct}\relax
\EndOfBibitem
\bibitem[Montoya and Haynes(2011)Montoya, and Haynes]{montoya2011periodic}
Montoya,~A.; Haynes,~B.~S. Periodic density functional study of Co$_3$O$_4$
  surfaces. \emph{Chem. Phys. Lett.} \textbf{2011}, \emph{502}, 63--68\relax
\mciteBstWouldAddEndPuncttrue
\mciteSetBstMidEndSepPunct{\mcitedefaultmidpunct}
{\mcitedefaultendpunct}{\mcitedefaultseppunct}\relax
\EndOfBibitem
\bibitem[Schwarz \latin{et~al.}(2018)Schwarz, Faisal, Mohr, Hohner, Werner, Xu,
  Skala, Tsud, Prince, Matolin, \latin{et~al.} others]{schwarz2018structure}
Schwarz,~M.; Faisal,~F.; Mohr,~S.; Hohner,~C.; Werner,~K.; Xu,~T.; Skala,~T.;
  Tsud,~N.; Prince,~K.~C.; Matolin,~V.; others Structure-dependent dissociation
  of water on cobalt oxide. \emph{J. Phys. Chem. Lett.} \textbf{2018},
  \emph{9}, 2763--2769\relax
\mciteBstWouldAddEndPuncttrue
\mciteSetBstMidEndSepPunct{\mcitedefaultmidpunct}
{\mcitedefaultendpunct}{\mcitedefaultseppunct}\relax
\EndOfBibitem
\bibitem[Budiyanto \latin{et~al.}(2023)Budiyanto, Ochoa-Hern{\'a}ndez, and
  T{\"u}ys{\"u}z]{budiyanto2023impact}
Budiyanto,~E.; Ochoa-Hern{\'a}ndez,~C.; T{\"u}ys{\"u}z,~H. Impact of Highly
  Concentrated Alkaline Treatment on Mesostructured Cobalt Oxide for the Oxygen
  Evolution Reaction. \emph{Adv. Sustain. Syst.} \textbf{2023}, \emph{7},
  2200499\relax
\mciteBstWouldAddEndPuncttrue
\mciteSetBstMidEndSepPunct{\mcitedefaultmidpunct}
{\mcitedefaultendpunct}{\mcitedefaultseppunct}\relax
\EndOfBibitem
\bibitem[Natarajan \latin{et~al.}(2021)Natarajan, Munirathinam, and
  Yang]{natarajan2021operando}
Natarajan,~K.; Munirathinam,~E.; Yang,~T.~C. Operando investigation of
  structural and chemical origin of \ce{Co3O4} stability in acid under oxygen
  evolution reaction. \emph{ACS Appl. Mater. Interfaces} \textbf{2021},
  \emph{13}, 27140--27148\relax
\mciteBstWouldAddEndPuncttrue
\mciteSetBstMidEndSepPunct{\mcitedefaultmidpunct}
{\mcitedefaultendpunct}{\mcitedefaultseppunct}\relax
\EndOfBibitem
\bibitem[Tran-Phu \latin{et~al.}(2022)Tran-Phu, Daiyan, Leverett, Fusco,
  Tadich, Di~Bernardo, Kiy, Truong, Zhang, Chen, \latin{et~al.}
  others]{tran2022understanding}
Tran-Phu,~T.; Daiyan,~R.; Leverett,~J.; Fusco,~Z.; Tadich,~A.; Di~Bernardo,~I.;
  Kiy,~A.; Truong,~T.~N.; Zhang,~Q.; Chen,~H.; others Understanding the
  activity and stability of flame-made \ce{Co3O4} spinels: A route towards the
  scalable production of highly performing OER electrocatalysts. \emph{Chem.
  Eng. J.} \textbf{2022}, \emph{429}, 132180\relax
\mciteBstWouldAddEndPuncttrue
\mciteSetBstMidEndSepPunct{\mcitedefaultmidpunct}
{\mcitedefaultendpunct}{\mcitedefaultseppunct}\relax
\EndOfBibitem
\bibitem[Qiu \latin{et~al.}(2024)Qiu, Maroun, Bouvier, Pacheco, Allongue,
  Wiegmann, Hendric~Scharf, de~Manuel-Gonzalez, Reikowski, Stettner,
  \latin{et~al.} others]{qiu2024operando}
Qiu,~C.; Maroun,~F.; Bouvier,~M.; Pacheco,~I.; Allongue,~P.; Wiegmann,~T.;
  Hendric~Scharf,~C.; de~Manuel-Gonzalez,~V.; Reikowski,~F.; Stettner,~J.;
  others Operando Surface X-Ray Diffraction Studies of Epitaxial \ce{Co3O4} and
  CoOOH Thin Films During Oxygen Evolution: pH Dependence. \emph{ChemCatChem}
  \textbf{2024}, \emph{16}, e202400988\relax
\mciteBstWouldAddEndPuncttrue
\mciteSetBstMidEndSepPunct{\mcitedefaultmidpunct}
{\mcitedefaultendpunct}{\mcitedefaultseppunct}\relax
\EndOfBibitem
\bibitem[Wiegmann \latin{et~al.}(2022)Wiegmann, Pacheco, Reikowski, Stettner,
  Qiu, Bouvier, Bertram, Faisal, Brummel, Libuda, \latin{et~al.}
  others]{wiegmann2022operando}
Wiegmann,~T.; Pacheco,~I.; Reikowski,~F.; Stettner,~J.; Qiu,~C.; Bouvier,~M.;
  Bertram,~M.; Faisal,~F.; Brummel,~O.; Libuda,~J.; others Operando
  identification of the reversible skin layer on \ce{Co3O4} as a
  three-dimensional reaction zone for oxygen evolution. \emph{ACS Catal.}
  \textbf{2022}, \emph{12}, 3256--3268\relax
\mciteBstWouldAddEndPuncttrue
\mciteSetBstMidEndSepPunct{\mcitedefaultmidpunct}
{\mcitedefaultendpunct}{\mcitedefaultseppunct}\relax
\EndOfBibitem
\bibitem[Reikowski \latin{et~al.}(2019)Reikowski, Maroun, Pacheco, Wiegmann,
  Allongue, Stettner, and Magnussen]{reikowski2019operando}
Reikowski,~F.; Maroun,~F.; Pacheco,~I.; Wiegmann,~T.; Allongue,~P.;
  Stettner,~J.; Magnussen,~O.~M. Operando surface X-ray diffraction studies of
  structurally defined \ce{Co3O4} and CoOOH thin films during oxygen evolution.
  \emph{ACS Catal.} \textbf{2019}, \emph{9}, 3811--3821\relax
\mciteBstWouldAddEndPuncttrue
\mciteSetBstMidEndSepPunct{\mcitedefaultmidpunct}
{\mcitedefaultendpunct}{\mcitedefaultseppunct}\relax
\EndOfBibitem
\bibitem[Haunold \latin{et~al.}(2025)Haunold, Anic, Genest, Rameshan, Roiaz,
  Li, Wicht, Knudsen, and Rupprechter]{haunold2025hydroxylation}
Haunold,~T.; Anic,~K.; Genest,~A.; Rameshan,~C.; Roiaz,~M.; Li,~H.; Wicht,~T.;
  Knudsen,~J.; Rupprechter,~G. Hydroxylation of an ultrathin \ce{Co3O4} (111)
  film on Ir (100) studied by in situ ambient pressure XPS and DFT. \emph{Surf.
  Sci.} \textbf{2025}, \emph{751}, 122618\relax
\mciteBstWouldAddEndPuncttrue
\mciteSetBstMidEndSepPunct{\mcitedefaultmidpunct}
{\mcitedefaultendpunct}{\mcitedefaultseppunct}\relax
\EndOfBibitem
\bibitem[Zhang \latin{et~al.}(2018)Zhang, Chen, Kamat, and
  Ptasinska]{zhang2018probing}
Zhang,~X.; Chen,~Y.-S.; Kamat,~P.~V.; Ptasinska,~S. Probing interfacial
  electrochemistry on a \ce{Co3O4} water oxidation catalyst using lab-based
  ambient pressure X-ray photoelectron spectroscopy. \emph{J. Phys. Chem. C}
  \textbf{2018}, \emph{122}, 13894--13901\relax
\mciteBstWouldAddEndPuncttrue
\mciteSetBstMidEndSepPunct{\mcitedefaultmidpunct}
{\mcitedefaultendpunct}{\mcitedefaultseppunct}\relax
\EndOfBibitem
\bibitem[Varhade \latin{et~al.}(2023)Varhade, Tetteh, Saddeler, Schumacher,
  Aiyappa, Bendt, Schulz, Andronescu, and Schuhmann]{varhade2023crystal}
Varhade,~S.; Tetteh,~E.~B.; Saddeler,~S.; Schumacher,~S.; Aiyappa,~H.~B.;
  Bendt,~G.; Schulz,~S.; Andronescu,~C.; Schuhmann,~W. Crystal Plane-Related
  Oxygen-Evolution Activity of Single Hexagonal \ce{Co3O4} Spinel Particles.
  \emph{Chem. Eur. J.} \textbf{2023}, \emph{29}, e202203474\relax
\mciteBstWouldAddEndPuncttrue
\mciteSetBstMidEndSepPunct{\mcitedefaultmidpunct}
{\mcitedefaultendpunct}{\mcitedefaultseppunct}\relax
\EndOfBibitem
\bibitem[Chen and Selloni(2012)Chen, and Selloni]{chen2012water}
Chen,~J.; Selloni,~A. Water adsorption and oxidation at the \ce{Co3O4} (110)
  surface. \emph{J. Phys. Chem. Lett.} \textbf{2012}, \emph{3},
  2808--2814\relax
\mciteBstWouldAddEndPuncttrue
\mciteSetBstMidEndSepPunct{\mcitedefaultmidpunct}
{\mcitedefaultendpunct}{\mcitedefaultseppunct}\relax
\EndOfBibitem
\bibitem[Kaptagay \latin{et~al.}(2015)Kaptagay, Inerbaev, Mastrikov, Kotomin,
  and Akilbekov]{kaptagay2015water}
Kaptagay,~G.; Inerbaev,~T.; Mastrikov,~Y.~A.; Kotomin,~E.; Akilbekov,~A. Water
  interaction with perfect and fluorine-doped \ce{Co3O4} (100) surface.
  \emph{Solid State Ionics} \textbf{2015}, \emph{277}, 77--82\relax
\mciteBstWouldAddEndPuncttrue
\mciteSetBstMidEndSepPunct{\mcitedefaultmidpunct}
{\mcitedefaultendpunct}{\mcitedefaultseppunct}\relax
\EndOfBibitem
\bibitem[Yan and Sautet(2019)Yan, and Sautet]{yan2019surface}
Yan,~G.; Sautet,~P. Surface structure of \ce{Co3O4} (111) under reactive
  gas-phase environments. \emph{ACS Catal.} \textbf{2019}, \emph{9},
  6380--6392\relax
\mciteBstWouldAddEndPuncttrue
\mciteSetBstMidEndSepPunct{\mcitedefaultmidpunct}
{\mcitedefaultendpunct}{\mcitedefaultseppunct}\relax
\EndOfBibitem
\bibitem[Yan \latin{et~al.}(2019)Yan, Wähler, Schuster, Schwarz, Hohner,
  Werner, Libuda, and Sautet]{yan2019water}
Yan,~G.; Wähler,~T.; Schuster,~R.; Schwarz,~M.; Hohner,~C.; Werner,~K.;
  Libuda,~J.; Sautet,~P. Water on oxide surfaces: a triaqua surface
  coordination complex on \ce{Co3O4} (111). \emph{J. Am. Chem. Soc.}
  \textbf{2019}, \emph{141}, 5623--5627\relax
\mciteBstWouldAddEndPuncttrue
\mciteSetBstMidEndSepPunct{\mcitedefaultmidpunct}
{\mcitedefaultendpunct}{\mcitedefaultseppunct}\relax
\EndOfBibitem
\bibitem[Huo \latin{et~al.}(2024)Huo, Lang, Song, Wang, Ren, Liao, Guo, and
  Chen]{huo2024dft}
Huo,~C.; Lang,~X.; Song,~G.; Wang,~Y.; Ren,~S.; Liao,~W.; Guo,~H.; Chen,~X. A
  DFT investigation on surface and defect modulation of the \ce{Co3O4} catalyst
  for efficient oxygen evolution reaction. \emph{Surf. Sci.} \textbf{2024},
  \emph{748}, 122544\relax
\mciteBstWouldAddEndPuncttrue
\mciteSetBstMidEndSepPunct{\mcitedefaultmidpunct}
{\mcitedefaultendpunct}{\mcitedefaultseppunct}\relax
\EndOfBibitem
\bibitem[Peng \latin{et~al.}(2021)Peng, Hajiyani, and
  Pentcheva]{peng2021influence}
Peng,~Y.; Hajiyani,~H.; Pentcheva,~R. Influence of Fe and Ni doping on the OER
  performance at the \ce{Co3O4} (001) surface: insights from DFT+U
  calculations. \emph{ACS Catal.} \textbf{2021}, \emph{11}, 5601--5613\relax
\mciteBstWouldAddEndPuncttrue
\mciteSetBstMidEndSepPunct{\mcitedefaultmidpunct}
{\mcitedefaultendpunct}{\mcitedefaultseppunct}\relax
\EndOfBibitem
\bibitem[Kox \latin{et~al.}(2020)Kox, Spohr, and Kenmoe]{kox2020impact}
Kox,~T.; Spohr,~E.; Kenmoe,~S. Impact of solvation on the structure and
  reactivity of the Co$_3$O$_4$(001)/H$_2$O interface: Insights from molecular
  dynamics simulations. \emph{Front. Energy Res.} \textbf{2020}, \emph{8},
  604799\relax
\mciteBstWouldAddEndPuncttrue
\mciteSetBstMidEndSepPunct{\mcitedefaultmidpunct}
{\mcitedefaultendpunct}{\mcitedefaultseppunct}\relax
\EndOfBibitem
\bibitem[Creazzo \latin{et~al.}(2019)Creazzo, Galimberti, Pezzotti, and
  Gaigeot]{creazzo2019dft}
Creazzo,~F.; Galimberti,~D.~R.; Pezzotti,~S.; Gaigeot,~M.-P. DFT-MD of the
  (110)-Co$_3$O$_4$ cobalt oxide semiconductor in contact with liquid water,
  preliminary chemical and physical insights into the electrochemical
  environment. \emph{J. Chem. Phys.} \textbf{2019}, \emph{150}\relax
\mciteBstWouldAddEndPuncttrue
\mciteSetBstMidEndSepPunct{\mcitedefaultmidpunct}
{\mcitedefaultendpunct}{\mcitedefaultseppunct}\relax
\EndOfBibitem
\bibitem[Kox and Kenmoe(2024)Kox, and Kenmoe]{kox2024co}
Kox,~T.; Kenmoe,~S. Co$_3$O$_4$ (111) surfaces in contact with water: molecular
  dynamics study of the surface chemistry and structure at room temperature.
  \emph{Dalton Trans.} \textbf{2024}, \emph{53}, 13184--13194\relax
\mciteBstWouldAddEndPuncttrue
\mciteSetBstMidEndSepPunct{\mcitedefaultmidpunct}
{\mcitedefaultendpunct}{\mcitedefaultseppunct}\relax
\EndOfBibitem
\bibitem[Behler(2016)]{P4885}
Behler,~J. Perspective: Machine Learning Potentials for Atomistic Simulations.
  \emph{J. Chem. Phys.} \textbf{2016}, \emph{145}, 170901\relax
\mciteBstWouldAddEndPuncttrue
\mciteSetBstMidEndSepPunct{\mcitedefaultmidpunct}
{\mcitedefaultendpunct}{\mcitedefaultseppunct}\relax
\EndOfBibitem
\bibitem[Deringer \latin{et~al.}(2019)Deringer, Caro, and Cs\'{a}nyi]{P5673}
Deringer,~V.~L.; Caro,~M.~A.; Cs\'{a}nyi,~G. Machine Learning Interatomic
  Potentials as Emerging Tools for Materials Science. \emph{Adv. Mater.}
  \textbf{2019}, \emph{31}, 1902765\relax
\mciteBstWouldAddEndPuncttrue
\mciteSetBstMidEndSepPunct{\mcitedefaultmidpunct}
{\mcitedefaultendpunct}{\mcitedefaultseppunct}\relax
\EndOfBibitem
\bibitem[Unke \latin{et~al.}(2021)Unke, Chmiela, Sauceda, Gastegger, Poltavsky,
  Sch{\"u}tt, Tkatchenko, and M{\"u}ller]{P6102}
Unke,~O.~T.; Chmiela,~S.; Sauceda,~H.~E.; Gastegger,~M.; Poltavsky,~I.;
  Sch{\"u}tt,~K.~T.; Tkatchenko,~A.; M{\"u}ller,~K.-R. Machine Learning Force
  Fields. \emph{Chem. Rev.} \textbf{2021}, \emph{121}, 10142--10186\relax
\mciteBstWouldAddEndPuncttrue
\mciteSetBstMidEndSepPunct{\mcitedefaultmidpunct}
{\mcitedefaultendpunct}{\mcitedefaultseppunct}\relax
\EndOfBibitem
\bibitem[Friederich \latin{et~al.}(2021)Friederich, H{\"a}se, Proppe, and
  Aspuru-Guzik]{P6112}
Friederich,~P.; H{\"a}se,~F.; Proppe,~J.; Aspuru-Guzik,~A. Machine-learned
  potentials for next-generation matter simulations. \emph{Nat. Mater}
  \textbf{2021}, \emph{20}, 750--761\relax
\mciteBstWouldAddEndPuncttrue
\mciteSetBstMidEndSepPunct{\mcitedefaultmidpunct}
{\mcitedefaultendpunct}{\mcitedefaultseppunct}\relax
\EndOfBibitem
\bibitem[Kocer \latin{et~al.}(2022)Kocer, Ko, and Behler]{P6131}
Kocer,~E.; Ko,~T.~W.; Behler,~J. Neural Network Potentials: A Concise Overview
  of Methods. \emph{Ann. Rev. Phys. Chem.} \textbf{2022}, \emph{73},
  163--186\relax
\mciteBstWouldAddEndPuncttrue
\mciteSetBstMidEndSepPunct{\mcitedefaultmidpunct}
{\mcitedefaultendpunct}{\mcitedefaultseppunct}\relax
\EndOfBibitem
\bibitem[Hou \latin{et~al.}(2024)Hou, Tian, and Meng]{P7143}
Hou,~P.; Tian,~Y.; Meng,~X. Improving Molecular-Dynamics Simulations for
  Solid--Liquid Interfaces with Machine-Learning Interatomic Potentials.
  \emph{Chem. Eur. J.} \textbf{2024}, \emph{30}, e202401373\relax
\mciteBstWouldAddEndPuncttrue
\mciteSetBstMidEndSepPunct{\mcitedefaultmidpunct}
{\mcitedefaultendpunct}{\mcitedefaultseppunct}\relax
\EndOfBibitem
\bibitem[Schran \latin{et~al.}(2021)Schran, Thiemann, Rowe, M{\"u}ller,
  Marsalek, and Michaelides]{schran2021machine}
Schran,~C.; Thiemann,~F.~L.; Rowe,~P.; M{\"u}ller,~E.~A.; Marsalek,~O.;
  Michaelides,~A. Machine learning potentials for complex aqueous systems made
  simple. \emph{Proc. Natl. Acad. Sci. U.S.A.} \textbf{2021}, \emph{118},
  e2110077118\relax
\mciteBstWouldAddEndPuncttrue
\mciteSetBstMidEndSepPunct{\mcitedefaultmidpunct}
{\mcitedefaultendpunct}{\mcitedefaultseppunct}\relax
\EndOfBibitem
\bibitem[Omranpour \latin{et~al.}(2024)Omranpour, Montero De~Hijes, Behler, and
  Dellago]{omranpour2024perspective}
Omranpour,~A.; Montero De~Hijes,~P.; Behler,~J.; Dellago,~C. Perspective:
  Atomistic simulations of water and aqueous systems with machine learning
  potentials. \emph{J. Chem. Phys.} \textbf{2024}, \emph{160}\relax
\mciteBstWouldAddEndPuncttrue
\mciteSetBstMidEndSepPunct{\mcitedefaultmidpunct}
{\mcitedefaultendpunct}{\mcitedefaultseppunct}\relax
\EndOfBibitem
\bibitem[Omranpour \latin{et~al.}(2025)Omranpour, Elsner, Lausch, and
  Behler]{omranpour2025machine}
Omranpour,~A.; Elsner,~J.; Lausch,~K.~N.; Behler,~J. Machine learning
  potentials for heterogeneous catalysis. \emph{ACS Catal.} \textbf{2025},
  \emph{15}, 1616--1634\relax
\mciteBstWouldAddEndPuncttrue
\mciteSetBstMidEndSepPunct{\mcitedefaultmidpunct}
{\mcitedefaultendpunct}{\mcitedefaultseppunct}\relax
\EndOfBibitem
\bibitem[Eckhoff \latin{et~al.}(2020)Eckhoff, Sch{\"o}newald, Risch, Volkert,
  Bl{\"o}chl, and Behler]{P5866}
Eckhoff,~M.; Sch{\"o}newald,~F.; Risch,~M.; Volkert,~C.~A.; Bl{\"o}chl,~P.~E.;
  Behler,~J. Closing the Gap between Theory and Experiment for Lithium
  Manganese Oxide Spinels Using a High-Dimensional Neural Network Potential.
  \emph{Phys. Rev. B} \textbf{2020}, \emph{102}, 174102\relax
\mciteBstWouldAddEndPuncttrue
\mciteSetBstMidEndSepPunct{\mcitedefaultmidpunct}
{\mcitedefaultendpunct}{\mcitedefaultseppunct}\relax
\EndOfBibitem
\bibitem[Eckhoff \latin{et~al.}(2020)Eckhoff, Lausch, Bl{\"o}chl, and
  Behler]{P5867}
Eckhoff,~M.; Lausch,~K.~N.; Bl{\"o}chl,~P.~E.; Behler,~J. Predicting Oxidation
  and Spin States by High-Dimensional Neural Networks: Applications to Lithium
  Manganese Oxide Spinels. \emph{J. Chem. Phys.} \textbf{2020}, \emph{153},
  164107\relax
\mciteBstWouldAddEndPuncttrue
\mciteSetBstMidEndSepPunct{\mcitedefaultmidpunct}
{\mcitedefaultendpunct}{\mcitedefaultseppunct}\relax
\EndOfBibitem
\bibitem[Eckhoff and Behler(2021)Eckhoff, and Behler]{P6141}
Eckhoff,~M.; Behler,~J. Insights into lithium manganese oxide-water interfaces
  using machine learning potentials. \emph{J. Chem. Phys.} \textbf{2021},
  \emph{155}, 244703\relax
\mciteBstWouldAddEndPuncttrue
\mciteSetBstMidEndSepPunct{\mcitedefaultmidpunct}
{\mcitedefaultendpunct}{\mcitedefaultseppunct}\relax
\EndOfBibitem
\bibitem[Schienbein and Blumberger(2022)Schienbein, and
  Blumberger]{schienbein2022nanosecond}
Schienbein,~P.; Blumberger,~J. Nanosecond solvation dynamics of the
  hematite/liquid water interface at hybrid DFT accuracy using committee neural
  network potentials. \emph{Phys. Chem. Chem. Phys.} \textbf{2022}, \emph{24},
  15365--15375\relax
\mciteBstWouldAddEndPuncttrue
\mciteSetBstMidEndSepPunct{\mcitedefaultmidpunct}
{\mcitedefaultendpunct}{\mcitedefaultseppunct}\relax
\EndOfBibitem
\bibitem[Romano \latin{et~al.}(2025)Romano, Montero~de Hijes, Meier, Kresse,
  Franchini, and Dellago]{romano2025structure}
Romano,~S.; Montero~de Hijes,~P.; Meier,~M.; Kresse,~G.; Franchini,~C.;
  Dellago,~C. Structure and dynamics of the magnetite (001)/water interface
  from molecular dynamics simulations based on a neural network potential.
  \emph{J. Chem. Theory Comput.} \textbf{2025}, \emph{21}, 1951--1960\relax
\mciteBstWouldAddEndPuncttrue
\mciteSetBstMidEndSepPunct{\mcitedefaultmidpunct}
{\mcitedefaultendpunct}{\mcitedefaultseppunct}\relax
\EndOfBibitem
\bibitem[Romano \latin{et~al.}(2024)Romano, Kaur, Zelenka, Hijes, Montero,
  Eder, Parkinson, Backus, and Dellago]{romano2024structure2}
Romano,~S.; Kaur,~H.; Zelenka,~M.; Hijes,~D.; Montero,~P.; Eder,~M.;
  Parkinson,~G.~S.; Backus,~E.~H.; Dellago,~C. Structure of the water/magnetite
  interface from sum frequency generation experiments and neural network based
  molecular dynamics simulations. \emph{arXiv preprint arXiv:2410.12717}
  \textbf{2024}, \relax
\mciteBstWouldAddEndPunctfalse
\mciteSetBstMidEndSepPunct{\mcitedefaultmidpunct}
{}{\mcitedefaultseppunct}\relax
\EndOfBibitem
\bibitem[Behler and Parrinello(2007)Behler, and
  Parrinello]{behler2007generalized}
Behler,~J.; Parrinello,~M. Generalized neural-network representation of
  high-dimensional potential-energy surfaces. \emph{Phys. Rev. Lett.}
  \textbf{2007}, \emph{98}, 146401\relax
\mciteBstWouldAddEndPuncttrue
\mciteSetBstMidEndSepPunct{\mcitedefaultmidpunct}
{\mcitedefaultendpunct}{\mcitedefaultseppunct}\relax
\EndOfBibitem
\bibitem[Behler(2021)]{behler2021four}
Behler,~J. Four generations of high-dimensional neural network potentials.
  \emph{Chem. Rev.} \textbf{2021}, \emph{121}, 10037--10072\relax
\mciteBstWouldAddEndPuncttrue
\mciteSetBstMidEndSepPunct{\mcitedefaultmidpunct}
{\mcitedefaultendpunct}{\mcitedefaultseppunct}\relax
\EndOfBibitem
\bibitem[Meredig \latin{et~al.}(2010)Meredig, Thompson, Hansen, Wolverton, and
  Van~de Walle]{meredig2010method}
Meredig,~B.; Thompson,~A.; Hansen,~H.; Wolverton,~C.; Van~de Walle,~A. Method
  for locating low-energy solutions within DFT+U. \emph{Phys. Rev. B}
  \textbf{2010}, \emph{82}, 195128\relax
\mciteBstWouldAddEndPuncttrue
\mciteSetBstMidEndSepPunct{\mcitedefaultmidpunct}
{\mcitedefaultendpunct}{\mcitedefaultseppunct}\relax
\EndOfBibitem
\bibitem[Bart{\'o}k \latin{et~al.}(2010)Bart{\'o}k, Payne, Kondor, and
  Cs{\'a}nyi]{bartok2010gaussian}
Bart{\'o}k,~A.~P.; Payne,~M.~C.; Kondor,~R.; Cs{\'a}nyi,~G. Gaussian
  approximation potentials: The accuracy of quantum mechanics, without the
  electrons. \emph{Phys. Rev. Lett.} \textbf{2010}, \emph{104}, 136403\relax
\mciteBstWouldAddEndPuncttrue
\mciteSetBstMidEndSepPunct{\mcitedefaultmidpunct}
{\mcitedefaultendpunct}{\mcitedefaultseppunct}\relax
\EndOfBibitem
\bibitem[Behler(2011)]{behler2011atom}
Behler,~J. Atom-centered symmetry functions for constructing high-dimensional
  neural network potentials. \emph{J. Chem. Phys.} \textbf{2011},
  \emph{134}\relax
\mciteBstWouldAddEndPuncttrue
\mciteSetBstMidEndSepPunct{\mcitedefaultmidpunct}
{\mcitedefaultendpunct}{\mcitedefaultseppunct}\relax
\EndOfBibitem
\bibitem[Kalman(1960)]{kalman1960new}
Kalman,~R.~E. A new approach to linear filtering and prediction problems.
  \emph{J. Basic Eng.} \textbf{1960}, \relax
\mciteBstWouldAddEndPunctfalse
\mciteSetBstMidEndSepPunct{\mcitedefaultmidpunct}
{}{\mcitedefaultseppunct}\relax
\EndOfBibitem
\bibitem[Blank and Brown(1994)Blank, and Brown]{blank1994adaptive}
Blank,~T.~B.; Brown,~S.~D. Adaptive, global, extended Kalman filters for
  training feedforward neural networks. \emph{J. Chemom.} \textbf{1994},
  \emph{8}, 391--407\relax
\mciteBstWouldAddEndPuncttrue
\mciteSetBstMidEndSepPunct{\mcitedefaultmidpunct}
{\mcitedefaultendpunct}{\mcitedefaultseppunct}\relax
\EndOfBibitem
\bibitem[Behler(2017)]{behler2017first}
Behler,~J. First principles neural network potentials for reactive simulations
  of large molecular and condensed systems. \emph{Angew. Chem. Int. Ed.}
  \textbf{2017}, \emph{56}, 12828--12840\relax
\mciteBstWouldAddEndPuncttrue
\mciteSetBstMidEndSepPunct{\mcitedefaultmidpunct}
{\mcitedefaultendpunct}{\mcitedefaultseppunct}\relax
\EndOfBibitem
\bibitem[Behler(2014)]{behler2014representing}
Behler,~J. Representing potential energy surfaces by high-dimensional neural
  network potentials. \emph{J. Condens. Matter Phys.} \textbf{2014}, \emph{26},
  183001\relax
\mciteBstWouldAddEndPuncttrue
\mciteSetBstMidEndSepPunct{\mcitedefaultmidpunct}
{\mcitedefaultendpunct}{\mcitedefaultseppunct}\relax
\EndOfBibitem
\bibitem[Behler(2015)]{behler2015constructing}
Behler,~J. Constructing high-dimensional neural network potentials: a tutorial
  review. \emph{Int. J. Quantum Chem} \textbf{2015}, \emph{115},
  1032--1050\relax
\mciteBstWouldAddEndPuncttrue
\mciteSetBstMidEndSepPunct{\mcitedefaultmidpunct}
{\mcitedefaultendpunct}{\mcitedefaultseppunct}\relax
\EndOfBibitem
\bibitem[Tokita and Behler(2023)Tokita, and Behler]{tokita2023train}
Tokita,~A.~M.; Behler,~J. How to train a neural network potential. \emph{J.
  Chem. Phys.} \textbf{2023}, \emph{159}\relax
\mciteBstWouldAddEndPuncttrue
\mciteSetBstMidEndSepPunct{\mcitedefaultmidpunct}
{\mcitedefaultendpunct}{\mcitedefaultseppunct}\relax
\EndOfBibitem
\bibitem[Kresse and Furthm{\"u}ller(1996)Kresse, and
  Furthm{\"u}ller]{kresse1996efficient}
Kresse,~G.; Furthm{\"u}ller,~J. Efficient iterative schemes for ab initio
  total-energy calculations using a plane-wave basis set. \emph{Phys. Rev. B}
  \textbf{1996}, \emph{54}, 11169\relax
\mciteBstWouldAddEndPuncttrue
\mciteSetBstMidEndSepPunct{\mcitedefaultmidpunct}
{\mcitedefaultendpunct}{\mcitedefaultseppunct}\relax
\EndOfBibitem
\bibitem[Kresse and Furthm{\"u}ller(1996)Kresse, and
  Furthm{\"u}ller]{kresse1996efficiency}
Kresse,~G.; Furthm{\"u}ller,~J. Efficiency of ab-initio total energy
  calculations for metals and semiconductors using a plane-wave basis set.
  \emph{Comput. Mater. Sci.} \textbf{1996}, \emph{6}, 15--50\relax
\mciteBstWouldAddEndPuncttrue
\mciteSetBstMidEndSepPunct{\mcitedefaultmidpunct}
{\mcitedefaultendpunct}{\mcitedefaultseppunct}\relax
\EndOfBibitem
\bibitem[Perdew \latin{et~al.}(1996)Perdew, Burke, and
  Ernzerhof]{perdew1996generalized}
Perdew,~J.~P.; Burke,~K.; Ernzerhof,~M. Generalized gradient approximation made
  simple. \emph{Phys. Rev. Lett.} \textbf{1996}, \emph{77}, 3865\relax
\mciteBstWouldAddEndPuncttrue
\mciteSetBstMidEndSepPunct{\mcitedefaultmidpunct}
{\mcitedefaultendpunct}{\mcitedefaultseppunct}\relax
\EndOfBibitem
\bibitem[Klime{\v{s}} \latin{et~al.}(2009)Klime{\v{s}}, Bowler, and
  Michaelides]{klimevs2009chemical}
Klime{\v{s}},~J.; Bowler,~D.~R.; Michaelides,~A. Chemical accuracy for the van
  der Waals density functional. \emph{J. Condens. Matter Phys.} \textbf{2009},
  \emph{22}, 022201\relax
\mciteBstWouldAddEndPuncttrue
\mciteSetBstMidEndSepPunct{\mcitedefaultmidpunct}
{\mcitedefaultendpunct}{\mcitedefaultseppunct}\relax
\EndOfBibitem
\bibitem[Klime{\v{s}} \latin{et~al.}(2011)Klime{\v{s}}, Bowler, and
  Michaelides]{klimevs2011van}
Klime{\v{s}},~J.; Bowler,~D.~R.; Michaelides,~A. Van der Waals density
  functionals applied to solids. \emph{Phys. Rev. B} \textbf{2011}, \emph{83},
  195131\relax
\mciteBstWouldAddEndPuncttrue
\mciteSetBstMidEndSepPunct{\mcitedefaultmidpunct}
{\mcitedefaultendpunct}{\mcitedefaultseppunct}\relax
\EndOfBibitem
\bibitem[Dudarev \latin{et~al.}(1998)Dudarev, Botton, Savrasov, Humphreys, and
  Sutton]{dudarev1998electron}
Dudarev,~S.~L.; Botton,~G.~A.; Savrasov,~S.~Y.; Humphreys,~C.; Sutton,~A.~P.
  Electron-energy-loss spectra and the structural stability of nickel oxide: An
  LSDA+U study. \emph{Phys. Rev. B} \textbf{1998}, \emph{57}, 1505\relax
\mciteBstWouldAddEndPuncttrue
\mciteSetBstMidEndSepPunct{\mcitedefaultmidpunct}
{\mcitedefaultendpunct}{\mcitedefaultseppunct}\relax
\EndOfBibitem
\bibitem[Bl{\"o}chl(1994)]{blochl1994projector}
Bl{\"o}chl,~P.~E. Projector augmented-wave method. \emph{Phys. Rev. B}
  \textbf{1994}, \emph{50}, 17953\relax
\mciteBstWouldAddEndPuncttrue
\mciteSetBstMidEndSepPunct{\mcitedefaultmidpunct}
{\mcitedefaultendpunct}{\mcitedefaultseppunct}\relax
\EndOfBibitem
\bibitem[Kresse and Joubert(1999)Kresse, and Joubert]{kresse1999ultrasoft}
Kresse,~G.; Joubert,~D. From ultrasoft pseudopotentials to the projector
  augmented-wave method. \emph{Phys. Rev. B} \textbf{1999}, \emph{59},
  1758\relax
\mciteBstWouldAddEndPuncttrue
\mciteSetBstMidEndSepPunct{\mcitedefaultmidpunct}
{\mcitedefaultendpunct}{\mcitedefaultseppunct}\relax
\EndOfBibitem
\bibitem[Schran \latin{et~al.}(2020)Schran, Brezina, and
  Marsalek]{schran2020committee}
Schran,~C.; Brezina,~K.; Marsalek,~O. Committee neural network potentials
  control generalization errors and enable active learning. \emph{J. Chem.
  Phys.} \textbf{2020}, \emph{153}, 104105\relax
\mciteBstWouldAddEndPuncttrue
\mciteSetBstMidEndSepPunct{\mcitedefaultmidpunct}
{\mcitedefaultendpunct}{\mcitedefaultseppunct}\relax
\EndOfBibitem
\bibitem[Omranpour and Behler(2024)Omranpour, and Behler]{omranpour2024high}
Omranpour,~A.; Behler,~J. A high-dimensional neural network potential for
  Co$_3$O$_4$. \emph{J. Phys.: Condens. Matter} \textbf{2024}, \emph{37},
  095701\relax
\mciteBstWouldAddEndPuncttrue
\mciteSetBstMidEndSepPunct{\mcitedefaultmidpunct}
{\mcitedefaultendpunct}{\mcitedefaultseppunct}\relax
\EndOfBibitem
\bibitem[Evans(1985)]{P2758}
Evans,~B.~L.,~D. J.;~Holian The Nose-Hoover thermostat. \emph{J. Chem. Phys.}
  \textbf{1985}, \emph{83}, 4069\relax
\mciteBstWouldAddEndPuncttrue
\mciteSetBstMidEndSepPunct{\mcitedefaultmidpunct}
{\mcitedefaultendpunct}{\mcitedefaultseppunct}\relax
\EndOfBibitem
\bibitem[Eckhoff and Behler(2021)Eckhoff, and Behler]{eckhoff2021high}
Eckhoff,~M.; Behler,~J. High-dimensional neural network potentials for magnetic
  systems using spin-dependent atom-centered symmetry functions. \emph{Npj
  Comput. Mater.} \textbf{2021}, \emph{7}, 170\relax
\mciteBstWouldAddEndPuncttrue
\mciteSetBstMidEndSepPunct{\mcitedefaultmidpunct}
{\mcitedefaultendpunct}{\mcitedefaultseppunct}\relax
\EndOfBibitem
\bibitem[Eckhoff and Behler(2019)Eckhoff, and Behler]{eckhoff2019molecular}
Eckhoff,~M.; Behler,~J. From molecular fragments to the bulk: development of a
  neural network potential for MOF-5. \emph{J. Chem. Theory Comput.}
  \textbf{2019}, \emph{15}, 3793--3809\relax
\mciteBstWouldAddEndPuncttrue
\mciteSetBstMidEndSepPunct{\mcitedefaultmidpunct}
{\mcitedefaultendpunct}{\mcitedefaultseppunct}\relax
\EndOfBibitem
\bibitem[Plimpton(1995)]{plimpton1995fast}
Plimpton,~S. Fast parallel algorithms for short-range molecular dynamics.
  \emph{J. Comput. Phys.} \textbf{1995}, \emph{117}, 1--19\relax
\mciteBstWouldAddEndPuncttrue
\mciteSetBstMidEndSepPunct{\mcitedefaultmidpunct}
{\mcitedefaultendpunct}{\mcitedefaultseppunct}\relax
\EndOfBibitem
\bibitem[Singraber \latin{et~al.}(2019)Singraber, Morawietz, Behler, and
  Dellago]{singraber2019parallel}
Singraber,~A.; Morawietz,~T.; Behler,~J.; Dellago,~C. Parallel multistream
  training of high-dimensional neural network potentials. \emph{J. Chem. Theory
  Comput.} \textbf{2019}, \emph{15}, 3075--3092\relax
\mciteBstWouldAddEndPuncttrue
\mciteSetBstMidEndSepPunct{\mcitedefaultmidpunct}
{\mcitedefaultendpunct}{\mcitedefaultseppunct}\relax
\EndOfBibitem
\bibitem[Swope \latin{et~al.}(1982)Swope, Andersen, Berens, and
  Wilson]{swope1982computer}
Swope,~W.~C.; Andersen,~H.~C.; Berens,~P.~H.; Wilson,~K.~R. A computer
  simulation method for the calculation of equilibrium constants for the
  formation of physical clusters of molecules: Application to small water
  clusters. \emph{J. Chem. Phys.} \textbf{1982}, \emph{76}, 637--649\relax
\mciteBstWouldAddEndPuncttrue
\mciteSetBstMidEndSepPunct{\mcitedefaultmidpunct}
{\mcitedefaultendpunct}{\mcitedefaultseppunct}\relax
\EndOfBibitem
\bibitem[Nos{\'e}(1984)]{nose1984molecular}
Nos{\'e},~S. A molecular dynamics method for simulations in the canonical
  ensemble. \emph{Mol. Phys.} \textbf{1984}, \emph{52}, 255--268\relax
\mciteBstWouldAddEndPuncttrue
\mciteSetBstMidEndSepPunct{\mcitedefaultmidpunct}
{\mcitedefaultendpunct}{\mcitedefaultseppunct}\relax
\EndOfBibitem
\bibitem[Hoover(1985)]{hoover1985canonical}
Hoover,~W.~G. Canonical dynamics: Equilibrium phase-space distributions.
  \emph{Phys. Rev. A} \textbf{1985}, \emph{31}, 1695\relax
\mciteBstWouldAddEndPuncttrue
\mciteSetBstMidEndSepPunct{\mcitedefaultmidpunct}
{\mcitedefaultendpunct}{\mcitedefaultseppunct}\relax
\EndOfBibitem
\bibitem[Quaranta \latin{et~al.}(2017)Quaranta, Hellström, and
  Behler]{quaranta2017proton}
Quaranta,~V.; Hellström,~M.; Behler,~J. Proton-transfer mechanisms at the
  water--ZnO interface: The role of presolvation. \emph{J. Phys. Chem. Lett.}
  \textbf{2017}, \emph{8}, 1476--1483\relax
\mciteBstWouldAddEndPuncttrue
\mciteSetBstMidEndSepPunct{\mcitedefaultmidpunct}
{\mcitedefaultendpunct}{\mcitedefaultseppunct}\relax
\EndOfBibitem
\bibitem[Henderson(2002)]{henderson2002interaction}
Henderson,~M.~A. The interaction of water with solid surfaces: fundamental
  aspects revisited. \emph{Surf. Sci. Rep.} \textbf{2002}, \emph{46},
  1--308\relax
\mciteBstWouldAddEndPuncttrue
\mciteSetBstMidEndSepPunct{\mcitedefaultmidpunct}
{\mcitedefaultendpunct}{\mcitedefaultseppunct}\relax
\EndOfBibitem
\bibitem[Björneholm \latin{et~al.}(2016)Björneholm, Hansen, Hodgson, Liu,
  Limmer, Michaelides, Pedevilla, Rossmeisl, Shen, Tocci, Tyrode, Walz, Werner,
  and Bluhm]{björneholm2016water}
Björneholm,~O.; Hansen,~M.~H.; Hodgson,~A.; Liu,~L.-M.; Limmer,~D.~T.;
  Michaelides,~A.; Pedevilla,~P.; Rossmeisl,~J.; Shen,~H.; Tocci,~G.;
  Tyrode,~E.; Walz,~M.-M.; Werner,~J.; Bluhm,~H. Water at interfaces.
  \emph{Chem. Rev.} \textbf{2016}, \emph{116}, 7698--7726\relax
\mciteBstWouldAddEndPuncttrue
\mciteSetBstMidEndSepPunct{\mcitedefaultmidpunct}
{\mcitedefaultendpunct}{\mcitedefaultseppunct}\relax
\EndOfBibitem
\bibitem[Davies(2016)]{davies2016role}
Davies,~P.~R. On the role of water in heterogeneous catalysis: A tribute to
  Professor M. Wyn Roberts. \emph{Top. Catal.} \textbf{2016}, \emph{59},
  671--677\relax
\mciteBstWouldAddEndPuncttrue
\mciteSetBstMidEndSepPunct{\mcitedefaultmidpunct}
{\mcitedefaultendpunct}{\mcitedefaultseppunct}\relax
\EndOfBibitem
\bibitem[Lin \latin{et~al.}(2021)Lin, Ge, Zhang, Wang, Xiao, and
  Ma]{lin2021heterogeneous}
Lin,~L.; Ge,~Y.; Zhang,~H.; Wang,~M.; Xiao,~D.; Ma,~D. Heterogeneous catalysis
  in water. \emph{JACS Au} \textbf{2021}, \emph{1}, 1834--1848\relax
\mciteBstWouldAddEndPuncttrue
\mciteSetBstMidEndSepPunct{\mcitedefaultmidpunct}
{\mcitedefaultendpunct}{\mcitedefaultseppunct}\relax
\EndOfBibitem
\bibitem[Stuchebrukhov and Hammes-Schiffer(2010)Stuchebrukhov, and
  Hammes-Schiffer]{stuchebrukhov2010theory}
Stuchebrukhov,~A.; Hammes-Schiffer,~S. Theory of coupled electron and proton
  transfer reactions. \emph{Chem. Rev.} \textbf{2010}, \emph{110},
  6939--6960\relax
\mciteBstWouldAddEndPuncttrue
\mciteSetBstMidEndSepPunct{\mcitedefaultmidpunct}
{\mcitedefaultendpunct}{\mcitedefaultseppunct}\relax
\EndOfBibitem
\bibitem[Agmon \latin{et~al.}(2016)Agmon, Bakker, Campen, Henchman, Pohl, Roke,
  Thamer, and Hassanali]{agmon2016protons}
Agmon,~N.; Bakker,~H.~J.; Campen,~R.~K.; Henchman,~R.~H.; Pohl,~P.; Roke,~S.;
  Thamer,~M.; Hassanali,~A. Protons and hydroxide ions in aqueous systems.
  \emph{Chem. Rev.} \textbf{2016}, \emph{116}, 7642--7672\relax
\mciteBstWouldAddEndPuncttrue
\mciteSetBstMidEndSepPunct{\mcitedefaultmidpunct}
{\mcitedefaultendpunct}{\mcitedefaultseppunct}\relax
\EndOfBibitem
\bibitem[Mu{\~n}oz-Santiburcio \latin{et~al.}(2018)Mu{\~n}oz-Santiburcio,
  Farnesi~Camellone, and Marx]{munoz2018solvation}
Mu{\~n}oz-Santiburcio,~D.; Farnesi~Camellone,~M.; Marx,~D. Solvation-induced
  changes in the mechanism of alcohol oxidation at gold/titania nanocatalysts
  in the aqueous phase versus gas phase. \emph{Angew. Chem.} \textbf{2018},
  \emph{130}, 3385--3389\relax
\mciteBstWouldAddEndPuncttrue
\mciteSetBstMidEndSepPunct{\mcitedefaultmidpunct}
{\mcitedefaultendpunct}{\mcitedefaultseppunct}\relax
\EndOfBibitem
\bibitem[Tran \latin{et~al.}(2016)Tran, Doan, Chandler, and
  Grabow]{tran2016water}
Tran,~H.-V.; Doan,~H.~A.; Chandler,~B.~D.; Grabow,~L.~C. Water-assisted oxygen
  activation during selective oxidation reactions. \emph{Curr. Opin. Chem.
  Eng.} \textbf{2016}, \emph{13}, 100--108\relax
\mciteBstWouldAddEndPuncttrue
\mciteSetBstMidEndSepPunct{\mcitedefaultmidpunct}
{\mcitedefaultendpunct}{\mcitedefaultseppunct}\relax
\EndOfBibitem
\end{mcitethebibliography}

\bibliographystyle{achemso}

\end{document}


\setstretch{1.0}
\title{Supplementary Information: Insights into the Structure and Dynamics of Water at Co$_3$O$_4$(001) Using a High-Dimensional Neural Network Potential}

\author{Amir Omranpour}
\email{\textcolor{black}{amir.omranpour@rub.de}}
\affiliation{Lehrstuhl f\"ur Theoretische Chemie II, Ruhr-Universit\"at Bochum, 44780 Bochum, Germany}
\affiliation{Research Center Chemical Sciences and Sustainability, Research Alliance Ruhr, 44780 Bochum, Germany}

\author{J\"{o}rg Behler}
\email{\textcolor{black}{joerg.behler@rub.de}}
\affiliation{Lehrstuhl f\"ur Theoretische Chemie II, Ruhr-Universit\"at Bochum, 44780 Bochum, Germany}
\affiliation{Research Center Chemical Sciences and Sustainability, Research Alliance Ruhr, 44780 Bochum, Germany}

\date{\today}

\maketitle

\section{Correlation Plots}\label{sec:correlation}

\begin{figure}[H]
\centering
\begin{subfigure}[b]{0.49\textwidth}
    \centering
    \includegraphics[width=\textwidth, trim= 0 0 40 0, clip=true]{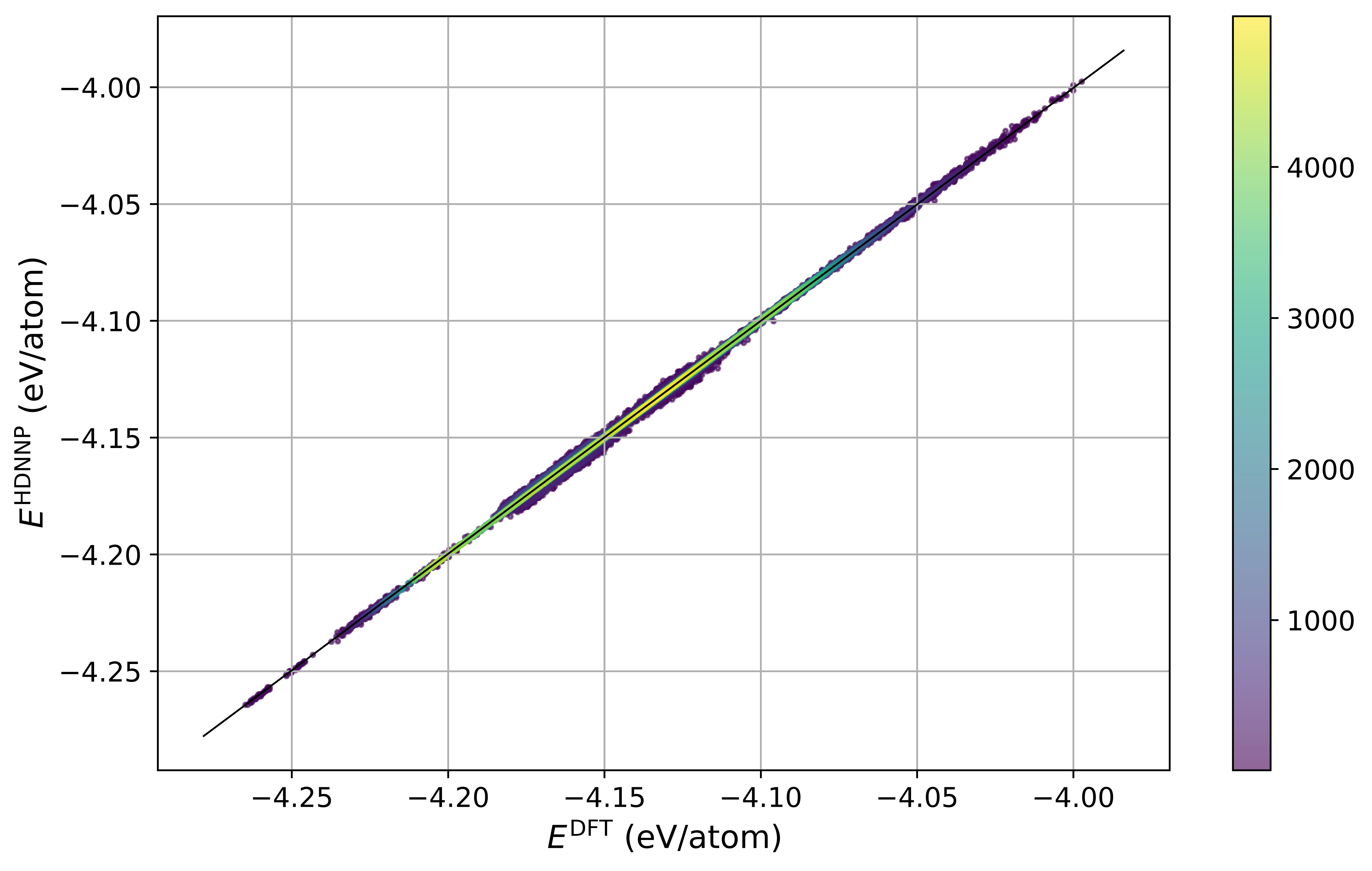}
    \caption{}
    \label{fig:dE_train}
\end{subfigure}
\hfill
\begin{subfigure}[b]{0.49\textwidth}
    \centering
    \includegraphics[width=\textwidth, trim= 0 0 40 0, clip=true]{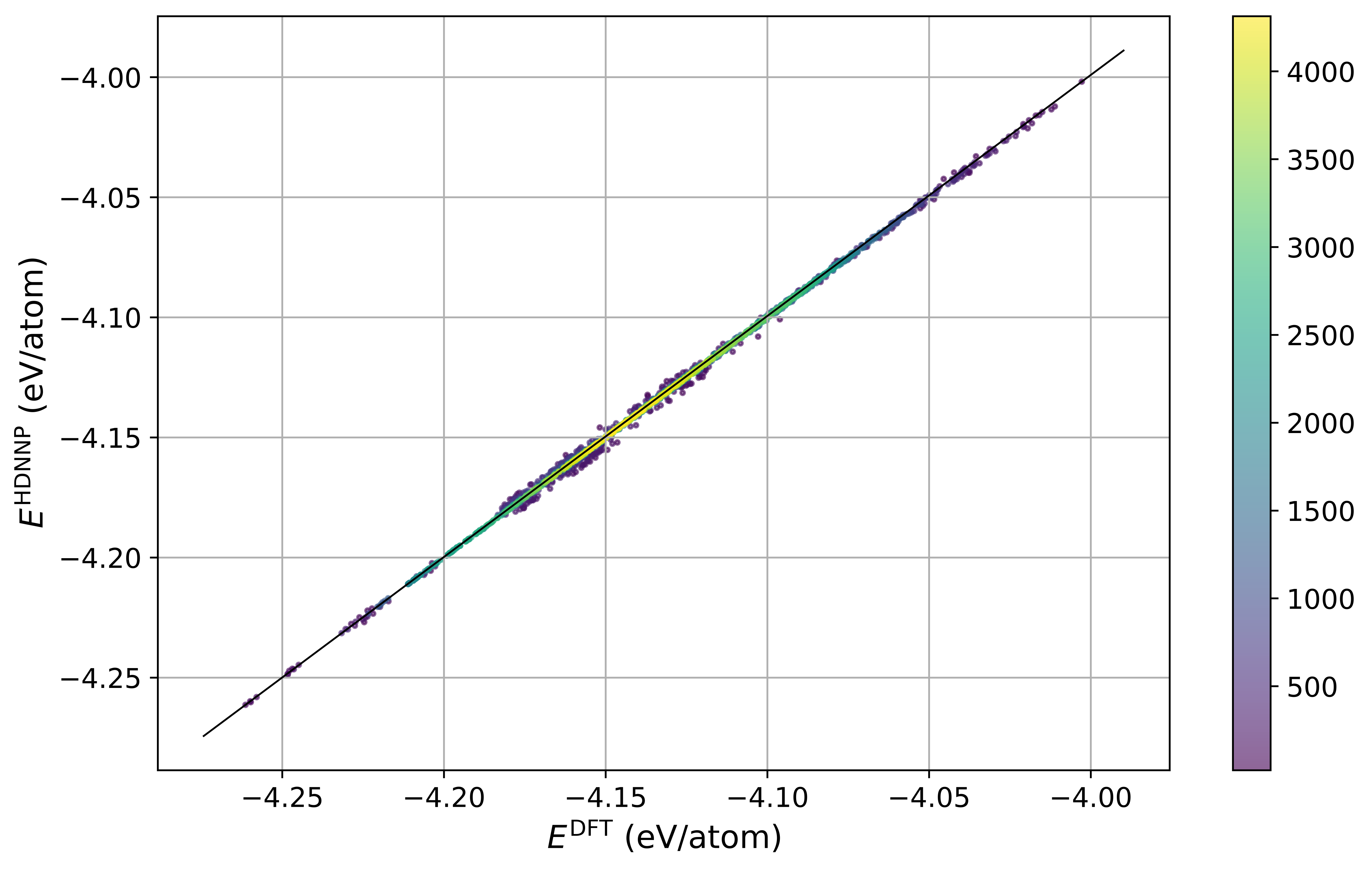}
    \caption{}
    \label{fig:dE_test}
\end{subfigure}
\caption{Energy correlation plots between the HDNNP predictions and DFT results for (a) the training dataset and (b) the testing dataset. The data points are color-coded based on their relative density, highlighting regions of higher data population.}
\label{fig:dE}
\end{figure}

\noindent Figure~\ref{fig:dE} reports energy–energy correlations between HDNNP predictions and DFT references for the training and test sets. Panel~\ref{fig:dE_train} shows the training data and panel~\ref{fig:dE_test} the test data. Points are colored by local point density (color bar on the right); brighter colors indicate regions where many configurations overlap. The black diagonal is the $y=x$ reference line and the overlaid fit line almost coincides with it across the full energy range. The cloud is narrow and uniform along $y=x$, with no drift at low or high energies, indicating minor bias and a stable slope close to unity. The test set exhibits the same behavior as the training set, confirming that the model reproduces DFT energies for unseen configurations and that no isolated outliers dominate the error.

\begin{figure}[H]
\centering
\begin{subfigure}[b]{0.49\textwidth}
    \centering
    \includegraphics[width=\textwidth, trim= 0 0 20 0, clip=true]{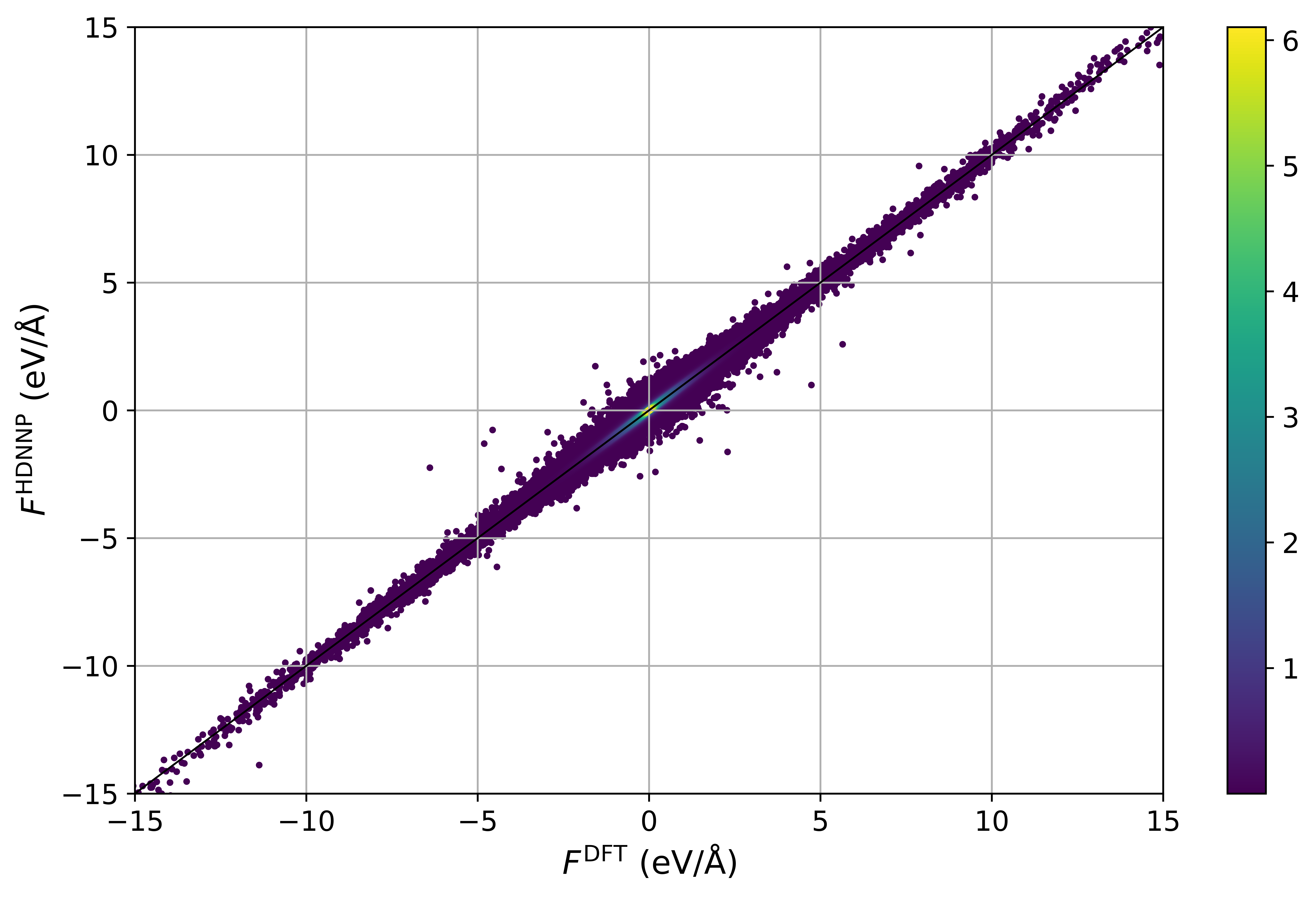}
    \caption{}
    \label{fig:dF_train}
\end{subfigure}
\hfill
\begin{subfigure}[b]{0.49\textwidth}
    \centering
    \includegraphics[width=\textwidth, trim= 0 0 30 0, clip=true]{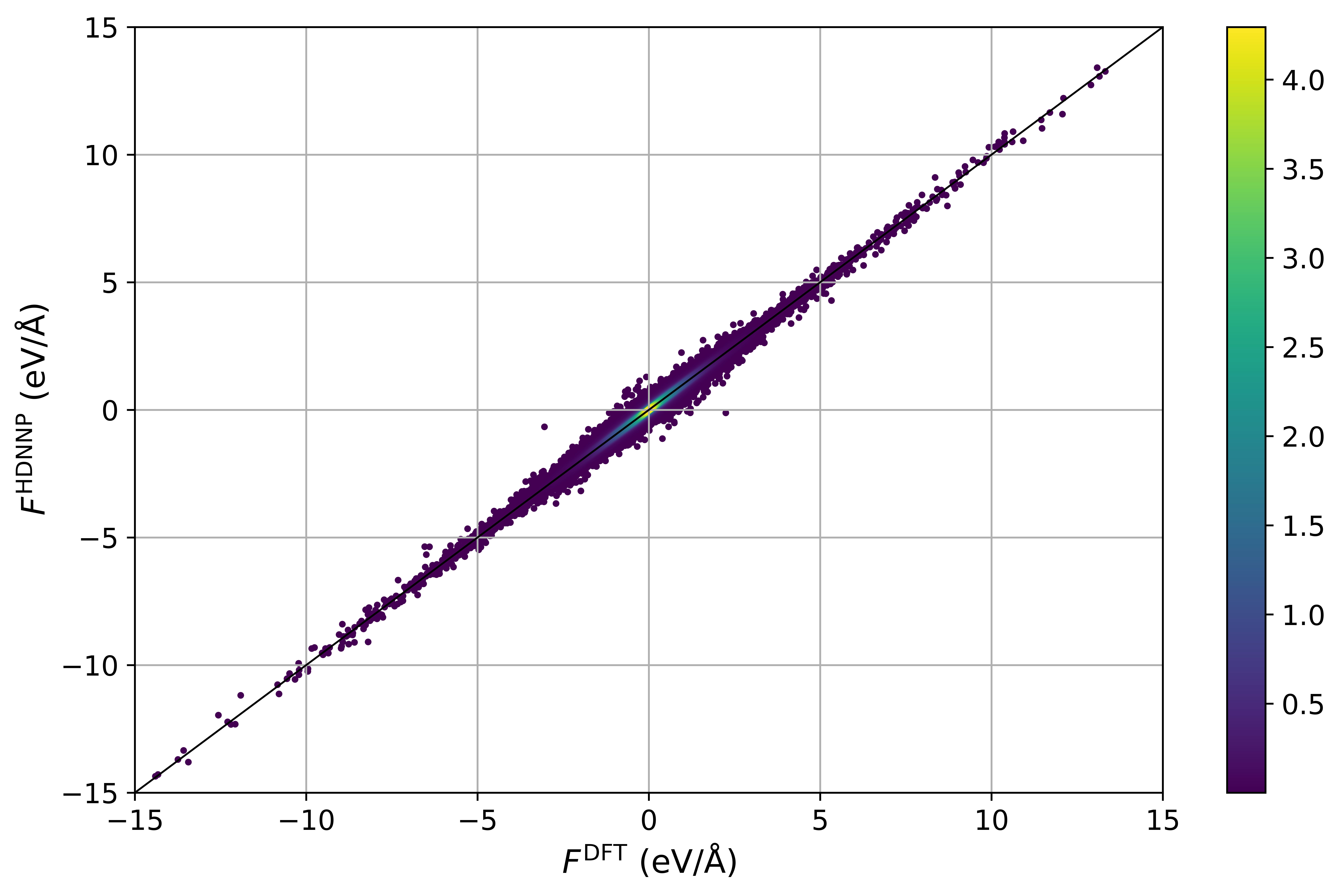}
    \caption{}
    \label{fig:dF_test}
\end{subfigure}
\caption{Force component correlation plots between the HDNNP predictions and DFT results for (a) the training dataset and (b) the testing dataset. The data points are color-coded based on their relative density, with brighter regions indicating areas of higher data population. Regions corresponding to high atomic forces originate from phases such as ice and water--gas interfaces included in the reference dataset (see the main text), but these are not visited in this work, which focuses exclusively on liquid water.}
\label{fig:dF}
\end{figure}

Figure~\ref{fig:dF} presents the force–force correlations. Panel~\ref{fig:dF_train} corresponds to the training set and panel~\ref{fig:dF_test} to the test set. As for the energies, the scatter is density colored; the most populated region is centered near small forces while the plot still includes large positive and negative force components up to the tails. The points cluster tightly around the $y=x$ line over the entire force range, and the fitted trend overlaps the unity line with only a small broadening at larger magnitudes. This indicates that the HDNNP captures both weak and strong forces without systematic bias, and that accuracy transfers from training to test data.

Taken together, the narrow, density–colored bands aligned with $y=x$ in Figures~\ref{fig:dE} and \ref{fig:dF} demonstrate that the HDNNP reproduces the DFT energy landscape and force field with high fidelity for both seen and unseen structures. It should be noted that the regions of high atomic forces originate from the associated phases of water, ice, and water--gas interfaces that were included in the reference dataset (see the main text). Such regions are not visited in this particular work, which deals exclusively with liquid water.

\section{Symmetry Functions}

\paragraph{Radial ACSFs.}
The radial symmetry functions are governed by the Gaussian width $\eta$ and an optional center shift $R_s$ that places sensitivity at specific neighbor shells. Table~\ref{tab:radial-acsf} lists the exact grids used for each center–neighbor pair (Co--Co, Co--O, Co--H, O--Co, O--O, O--H, H--Co, H--O, H--H). The unshifted sets ($R_s=0$) span $\eta \approx 10^{-3}$ to $6\times10^{-1}$~\AA$^{-2}$, providing resolution from broad to sharp radial features. In addition, a small group of $R_s$-shifted Gaussians with $\eta=0.0200$~\AA$^{-2}$ and $R_s\in\{1.0,\,1.5,\,2.0,\,3.0,\,4.5\}$~\AA\ is included to target first and second coordination shells at the oxide–water interface. All radial functions employ a cosine cutoff with $R_c=6.35$~\AA\ (corresponding to $12$~bohr).

\paragraph{Angular ACSFs.}
The angular symmetry functions encode the relative orientation of atomic triplets around each center atom. As summarized in Table~\ref{tab:angular-acsf}, we use the standard grid with $\eta=0$~\AA$^{-2}$, $\lambda\in\{+1,-1\}$ to set the sign of the angle term, and $\zeta\in\{1,2,4,16\}$ to control the angular sensitivity. This grid is applied uniformly to the unordered neighbor pairs \{Co--Co, O--O, Co--O, Co--H, O--H, H--H\} for each center species (Co, O, H), yielding a balanced set of purely angular descriptors under the same cosine cutoff $R_c=6.35$~\AA. Together with the radial channels, this basis resolves both the short-range coordination and the orientational preferences that characterize the Co$_3$O$_4$(001)--water interface.

\begin{table*}[ht]
\centering
\renewcommand{\arraystretch}{1.2}
\caption{Radial $G^{(2)}$ symmetry functions used. Columns list the center species, neighbor pair, the set of $\eta$ values (in \AA$^{-2}$), the Gaussian shift $R_s$ (in \AA), and cutoff $R_c$ (in \AA). All entries use a cosine cutoff with $R_c=6.35$~\AA\ ($12$ bohr)}
\label{tab:radial-acsf}
\small
\begin{tabular}{lllll}
\hline
Center & Pair & $\eta$ (\AA$^{-2}$) & $R_s$ (\AA) & $R_c$ (\AA) \\ \hline
Co & Co--Co & $\{0.0013,\,0.0032,\,0.0080,\,0.0200,\,0.0500,\,0.1000,\,0.2500\}$ & $0.0$ & $6.35$ \\ \hline
Co & Co--O  & $\{0.0020,\,0.0051,\,0.0128,\,0.0320,\,0.0800,\,0.1200,\,0.3000\}$ & $0.0$ & $6.35$ \\ \hline
Co & Co--O  & $\{0.0200\}$ & $\{1.5,\,3.0,\,4.5\}$ & $6.35$ \\ \hline
Co & Co--H  & $\{0.0025,\,0.0064,\,0.0160,\,0.0400,\,0.1000,\,0.1500,\,0.3500\}$ & $0.0$ & $6.35$ \\ \hline
O  & O--Co  & $\{0.0020,\,0.0051,\,0.0128,\,0.0320,\,0.0800,\,0.1200,\,0.3000\}$ & $0.0$ & $6.35$ \\ \hline
O  & O--Co  & $\{0.0200\}$ & $\{1.5,\,3.0,\,4.5\}$ & $6.35$ \\ \hline
O  & O--O   & $\{0.0013,\,0.0032,\,0.0080,\,0.0200,\,0.0500,\,0.1000,\,0.2500\}$ & $0.0$ & $6.35$ \\ \hline
O  & O--H   & $\{0.0030,\,0.0076,\,0.0190,\,0.0480,\,0.1200,\,0.2500,\,0.5000\}$ & $0.0$ & $6.35$ \\ \hline
O  & O--H   & $\{0.0200\}$ & $\{1.0,\,2.0,\,3.0\}$ & $6.35$ \\ \hline
H  & H--Co  & $\{0.0025,\,0.0064,\,0.0160,\,0.0400,\,0.1000,\,0.1500,\,0.3500\}$ & $0.0$ & $6.35$ \\ \hline
H  & H--O   & $\{0.0030,\,0.0076,\,0.0190,\,0.0480,\,0.1200,\,0.2500,\,0.5000\}$ & $0.0$ & $6.35$ \\ \hline
H  & H--O   & $\{0.0200\}$ & $\{1.0,\,2.0,\,3.0\}$ & $6.35$ \\ \hline
H  & H--H   & $\{0.0038,\,0.0096,\,0.0240,\,0.0600,\,0.1500,\,0.3000,\,0.6000\}$ & $0.0$ & $6.35$ \\ \hline
\end{tabular}
\end{table*}

\begin{table*}[ht]
\centering
\renewcommand{\arraystretch}{1.2}
\caption{Angular $G^{(4)}$ symmetry functions used. For all rows, $\eta=0$ \AA$^{-2}$, $\lambda\in\{+1,-1\}$, $\xi\in\{1,2,4,16\}$, and a cosine cutoff $R_c=6.35$ \AA\ ($12$ bohr)}
\label{tab:angular-acsf}
\small
\begin{tabular}{llllll}
\hline
Center & Pair & $\eta$ (\AA$^{-2}$) & $\lambda$ & $\xi$ & $R_c$ (\AA) \\ \hline
Co & Co--Co & $0$ & $\{+1,-1\}$ & $\{1,2,4,16\}$ & $6.35$ \\ \hline
Co & O--O   & $0$ & $\{+1,-1\}$ & $\{1,2,4,16\}$ & $6.35$ \\ \hline
Co & Co--O  & $0$ & $\{+1,-1\}$ & $\{1,2,4,16\}$ & $6.35$ \\ \hline
Co & Co--H  & $0$ & $\{+1,-1\}$ & $\{1,2,4,16\}$ & $6.35$ \\ \hline
Co & O--H   & $0$ & $\{+1,-1\}$ & $\{1,2,4,16\}$ & $6.35$ \\ \hline
Co & H--H   & $0$ & $\{+1,-1\}$ & $\{1,2,4,16\}$ & $6.35$ \\ \hline
O  & Co--Co & $0$ & $\{+1,-1\}$ & $\{1,2,4,16\}$ & $6.35$ \\ \hline
O  & O--O   & $0$ & $\{+1,-1\}$ & $\{1,2,4,16\}$ & $6.35$ \\ \hline
O  & Co--O  & $0$ & $\{+1,-1\}$ & $\{1,2,4,16\}$ & $6.35$ \\ \hline
O  & Co--H  & $0$ & $\{+1,-1\}$ & $\{1,2,4,16\}$ & $6.35$ \\ \hline
O  & O--H   & $0$ & $\{+1,-1\}$ & $\{1,2,4,16\}$ & $6.35$ \\ \hline
O  & H--H   & $0$ & $\{+1,-1\}$ & $\{1,2,4,16\}$ & $6.35$ \\ \hline
H  & Co--Co & $0$ & $\{+1,-1\}$ & $\{1,2,4,16\}$ & $6.35$ \\ \hline
H  & O--O   & $0$ & $\{+1,-1\}$ & $\{1,2,4,16\}$ & $6.35$ \\ \hline
H  & Co--O  & $0$ & $\{+1,-1\}$ & $\{1,2,4,16\}$ & $6.35$ \\ \hline
H  & Co--H  & $0$ & $\{+1,-1\}$ & $\{1,2,4,16\}$ & $6.35$ \\ \hline
H  & O--H   & $0$ & $\{+1,-1\}$ & $\{1,2,4,16\}$ & $6.35$ \\ \hline
H  & H--H   & $0$ & $\{+1,-1\}$ & $\{1,2,4,16\}$ & $6.35$ \\ \hline
\end{tabular}
\end{table*}

\newpage
\clearpage

\section{RuNNer Settings}

\noindent Table~\ref{tab:S4} presents the key settings used in the RuNNer input file for constructing the high-dimensional neural network potential for the  Co$_3$O$_4$(001)--Water Interfaces.

\begin{table}[H]
\centering
\renewcommand{\arraystretch}{1.3}
\caption{Settings in the RuNNer input file for constructing the HDNNP (specification of the ACSFs are left out).}
\label{tab:S4}
\begin{tabular}{ll}
\hline
\textbf{Setting} & \textbf{Value} \\ \hline
nn\_type\_short & 1 \\ \hline
random\_number\_type & 5 \\ \hline
random\_seed & 3000000000 \\ \hline
number\_of\_elements & 3 \\ \hline
elements & Co O H \\ \hline
cutoff\_type & 1 \\ \hline
use\_short\_nn & \\ \hline
global\_hidden\_layers\_short & 2 \\ \hline
global\_nodes\_short & 25 20 \\ \hline
global\_activation\_short & t t l \\ \hline
test\_fraction & 0.05 \\ \hline
epochs & 30 \\ \hline
points\_in\_memory & 1000 \\ \hline
mix\_all\_points & \\ \hline
scale\_symmetry\_functions & \\ \hline
center\_symmetry\_functions & \\ \hline
fitting\_unit & eV \\ \hline
precondition\_weights & \\ \hline
use\_short\_forces & \\ \hline
optmode\_short\_energy & 1 \\ \hline
optmode\_short\_force & 1 \\ \hline
kalman\_lambda\_short & 0.98 \\ \hline
kalman\_nue\_short & 0.9987 \\ \hline
short\_energy\_fraction & 1.0 \\ \hline
short\_force\_fraction & 0.1 \\ \hline
weights\_min & -1.0 \\ \hline
weights\_max & 1.0 \\ \hline
nguyen\_widrow\_weights\_short & \\ \hline
short\_force\_error\_threshold & 1.0 \\ \hline
\end{tabular}
\end{table}
